\documentclass[11pt]{amsart}
\usepackage{amsmath}
\usepackage{amsthm}
\usepackage{amsfonts}
\usepackage{amssymb}

\theoremstyle{definition}

\theoremstyle{remark}

\newcommand{\sst}{\scriptscriptstyle}
\newcommand{\dst}{\displaystyle}

\newcommand{\beq}{\begin{equation}}
\newcommand{\eeq}{\end{equation}}

\newcommand{\D}{\Delta}
\newcommand{\pa}{\partial}
\newcommand{\ot}{\otimes}
\newcommand{\ra}{\rightarrow}

\newcommand{\ti}{\times}

\newcommand{\fr}[2]{{\textstyle \frac{#1}{#2} }}

\newcommand{\fsl}{{\mathfrak s}{\mathfrak l}}
\newcommand{\hfsl}{\widehat{\fsl}}

\newcommand{\al}{\alpha}
\newcommand{\be}{\beta}
\newcommand{\ga}{\gamma}
\newcommand{\Ga}{\Gamma}
\newcommand{\de}{\delta}

\newcommand{\ep}{\epsilon}
\newcommand{\la}{\lambda}

\newcommand{\Om}{\Omega}
\newcommand{\si}{\sigma}
\newcommand{\up}{\Upsilon}
\newcommand{\vf}{\varphi}

\newcommand{\bJ}{\bar{J}}
\newcommand{\bI}{\bar{I}}
\newcommand{\bT}{\bar{T}}
\newcommand{\bL}{\bar{L}}
\newcommand{\ba}{\bar{a}}

\newcommand{\bk}{\bar{k}}
\newcommand{\bq}{\bar{q}}
\newcommand{\bn}{\bar{n}}
\newcommand{\bu}{\bar{u}}
\newcommand{\bv}{\Bar{v}}
\newcommand{\bw}{\bar{w}}
\newcommand{\bx}{\bar{x}}

\newcommand{\bz}{\bar{z}}
\newcommand{\bmu}{\bar{\mu}}

\newcommand{\CA}{{\mathcal A}}
\newcommand{\CB}{{\mathcal B}}
\newcommand{\CC}{{\mathcal C}}
\newcommand{\CD}{{\mathcal D}}

\newcommand{\CF}{{\mathcal F}}
\newcommand{\CG}{{\mathcal G}}
\newcommand{\CH}{{\mathcal H}}
\newcommand{\CI}{{\mathcal I}}
\newcommand{\CJ}{{\mathcal J}}

\newcommand{\CN}{{\mathcal N}}
\newcommand{\CO}{{\mathcal O}}  
\newcommand{\CP}{{\mathcal P}}  
  
\newcommand{\CR}{{\mathcal R}}
\newcommand{\CS}{{\mathcal S}}

\newcommand{\CU}{{\mathcal U}}
\newcommand{\CV}{{\mathcal V}}

\newcommand{\CZ}{{\mathcal Z}}
\newcommand{\fg}{{\mathfrak g}}
\newcommand{\fb}{{\mathfrak b}}
\newcommand{\fn}{{\mathfrak n}}

\newcommand{\hfg}{\hat{{\mathfrak g}}}

\newcommand{\BR}{{\mathbb R}}

\newcommand{\BC}{{\mathbb C}}

\newcommand{\BZ}{{\mathbb Z}}

\newcommand{\hp}{$H_3^+$}

\DeclareMathOperator{\Tr}{Tr}
\newcommand{\CGL}{{\mathcal G}_{}^{\sst L}{}}

\newcommand{\rf}[1]{(\ref{#1})}
\newcommand{\aufz}
{\begin{list}{$\bullet$}{\topsep0cm \itemsep0cm \parsep0cm}}
\newcommand{\eaufz}{\end{list}}
\newcounter{num}
\newcommand{\remlst}{\begin{list}
{(\arabic{num})}{\usecounter{num}\topsep0cm \itemsep0cm \parsep0cm}}
\begin{document}
\thispagestyle{empty}
\hspace*{\fill} LPM-97/03\\
\hspace*{\fill} hep-th/9712256\\[.5cm]
\title{On structure constants and fusion rules in the
$SL(2,\BC)/SU(2)$ WZNW model
\footnote{Work supported by the 
European Union under contract FMRX-CT96-0012}}
\author{\sc J. Teschner}
\address{Laboratoire de Physique Math\'ematique,
Universit\'{e} Montpellier II,
Pl. E. Bataillon, 34095 Montpellier, France}
\email{teschner@lpm.univ-montp2.fr}

\begin{abstract}
A closed formula for the structure constants in the SL(2,C)/SU(2) 
WZNW model is derived by a method previously used in Liouville theory. 
With the help of a reflection amplitude that follows from the structure 
constants one obtains a proposal for the fusion rules from canonical 
quantization. Taken together these pieces of information allow an unambigous 
definition of any genus zero n-point function.
\end{abstract}

\maketitle

\section{Introduction}

Little is known about conformal field theories that have continuous families
of primary fields if they are not related to free field theories in a simple
way. As the existence of continuous spectra of representations is connected
with noncompactness of zero-mode configuration space, one may call such 
theories non-compact CFT's.
But such theories promise to have a multitude of interesting 
applications and connections to other branches: Let me mention 
various quantum gravity models, connections with massive integrable theories
via perturbed CFT, connections with integrable models such as 
(non-compact versions of) Gaudin, Calogero-Moser, Hitchin etc.. 

Non-compact conformal field theories are expected to be
in many respects qualitatively different
from the well studied rational or compact CFT's:
The representations of the current algebra that the
primary field correspond to will in general have no highest or lowest
weight vector (so also no singular vectors). 
The fusion rules are therefore not determined algebraically
but rather analytically. One expects that the operator product expansions 
of primary fields will involve integrals over continuous sets of operators.
Non-vanishing of
three point functions does not imply that any one of the three operators
actually appears in the OPE of the other two.

The aim of the present work is to obtain some exact results for
one of the simplest examples of a noncompact CFT, the WZNW model 
corresponding to the coset $H_3^+=SL(2,\BC)/SU(2)$. 
It will become clear that the \hp\ WZNW model itself is not a good physical 
theory: It is not unitary. 
There are two main reasons for studying it nevertheless:

One important physical motivation 
for studying the \hp\ WZNW model 
comes from its relation to the euclidean black hole CFT, 
which is expected to be unitary (cf. discussion in \cite{Ga}). The point is
that the $\fsl_2$-current algebra symmetries are explicit in the 
\hp\ WZNW model and therefore help to construct primary fields and correlation
functions. Many of the results obtained here can be carried over
to the euclidean black hole CFT. There is also an intimate
relation to Liouville theory, which will be discussed further. 

The second reason is that it seems to be (besides Liouville) one of the 
simplest noncompact CFT's to study. It is therefore a natural starting
point for the developement of methods for the investigation of noncompact
CFT's, just as the study of representation theory of $SL(2,\BC)$ and
$SL(2,\BR)$ by Gelfand, Naimark and Bargmann was the
starting point for the representation theory of noncomapct groups 
in general. 

The present paper is part of a series of three papers on that subject. 
The first one \cite{T1} studies the semiclassical or mini-superspace limit
in detail. The theory essentially reduces to harmonic analysis on the 
symmetric space \hp, so a rather complete and rigorous treatment is possible
in this limit. This nicely illustrates how the new qualitative features
of the \hp\ WZNW model compared to compact CFT's can be understood from 
the point of view of harmonic analysis. 

The second one \cite{T2}
is devoted to certain mathematical aspects of the 
relevant current algebra representations and the construction of conformal 
blocks. It treats the cases of current algebra representations induced
from $SL(2,\BR)$, $SL(2,\BC)$ and $SU(2)$ zero mode representations in
a uniform manner and therefore lays some rigorous ground not only for the   
present work but also for a forthcoming treatment of the $SL(2,\BR)$
WZNW model.

The contents of the present paper may be summarized as follows:
The second section is a scetch of the canonical quantization of the
model. The current algebra symmetry is established and the stage is set
for the later discussion of fusion rules. The spectrum of representations
is discussed, based on the result of \cite{Ga} on the partition function.

Since the relevant representations are neither highest nor lowest weight 
representations one needs a generalization of the bootstrap approach
\cite{BPZ} that will be introduced in the following third section. It may
be considered as ``affinization'' of $SL(2,\BC)$ representation theory.

The following fourth section describes a derivation of the structure 
constants resp. three point functions. The method was previously used
for Liouville theory in \cite{TL}. It is based on the 
consideration of four point functions with one degenerate field which
satsify additional differential equations. Assuming crossing symmetry 
for these four point functions leads to functional relations for
the structure constants of three generic primary fields which have 
a unique solution for irrational central charge.  

Finally, the fifth section is devoted to a discussion of the issue of fusion 
rules. The results of \cite{Ga} suggests that any normalizable state
can be expanded in terms of states from irreducible representations $\CP_j$
of the current algebra, where $j=-1/2+i\rho$, $\rho\in\BR$ and the $\CP_j$ 
are induced from principal series representations of $SL(2,\BC)$.
The canonical normalization of states involves integration over
zero modes. A simple condition for normalizability is 
found by considering the zero mode asymptotics of the state
created by action of primary field $\Phi^{j_2}$ on a primary state 
$\Psi^{j_1}$. This defines a certain range of values for $j_2$, $j_1$ for which
the fusion rules are simply given by expansion over all of the spectrum.
The fusion rules for more general values of $j_1,j_2$ can be
obtained by analytic continuation.

 {\sc Acknowledgments}

I would like to thank A.B. and Al.B. Zamolodchikov for valuable discussions. 
They independently found some crucial results such as the structure constants.

\section{Canonical Quantization, Symmetries}

The coset 
$H_3^+\equiv SL(2,\BC)/SU(2)$ is the set of all hermitian two-by-two matrices
with determinant one. A convenient global coordinate system for \hp is 
provided by the parametrization
\begin{equation}
h=\begin{pmatrix} 1 & u \\ 0 & 1 \end{pmatrix}
\begin{pmatrix} e^{\phi} & 0 \\ 0 & e^{-\phi} \end{pmatrix}
\begin{pmatrix} 1 & 0 \\ \bu & 1 \end{pmatrix}
\end{equation}
The starting point of the present discussion will be the following 
action
\begin{equation} \label{action}
S=\frac{1}{\pi} \int_{-\infty}^{\infty}d\tau \int_{0}^{2\pi} d\si
\bigl( \pa_+\phi\pa_-\phi+\pa_+u\pa_-\bu e^{-2\phi}\bigr),
\end{equation}
where $\tau$ and $\si$ denote the time and (periodic) space variables 
respectively. This action may be obtained \cite{Ga} from a $SL(2,\BC)$-WZNW
model by gauging the $SU(2)$ subgroup in the Lagrangian formalism. Anyway,
action \rf{action} defines a sigma-model with target \hp that will be shown 
to possess conformal invariance via canonical quantization.

\subsection{Canonical quantization}

Introducing the canonical momenta as
\begin{equation}
 \Pi_{\phi}=2\pi\dot{\phi} \qquad \Pi_{u}=\pi(\Dot{\bu}-\bu')e^{-2\phi}
\qquad \Pi_{\bu}=\pi(\Dot{u}+u')e^{-2\phi} 
\end{equation}
leads to the following expression for the Hamiltonian 
\begin{equation}
H_{cl}= 
\frac{1}{2\pi}\int_{0}^{2\pi}d\si \Bigl(\fr{1}{2}(\Pi_{\phi}^2+(\phi')^2)
+2 \Pi_u\Pi_{\bu}e^{2\phi}
+\Pi_{\bu}\bu'-\Pi_u u' \Bigr)
\end{equation}
The Hamiltonian system thereby defined is quantized by postulating 
canonical commutation relations at time $\tau=0$:
\begin{equation}\label{canon}\begin{aligned}
{[}u(\si),\Pi_u(\si'){]}=2\pi ib^2\de(\si-\si')\\
[\bu(\si),\Pi_{\bu}(\si')]=2\pi ib^2\de(\si-\si')
\end{aligned}\qquad
[\phi(\si),\Pi_{\phi}(\si')]=2\pi ib^2\de(\si-\si'),
\end{equation}
all other commutators vanishing. Planck's constant $\hbar$ was for later
convenience written as $\hbar=b^2$. In the description of the quantum theory 
the rescaled fields $\vf=b^{-1}\phi$, $\Pi_{\vf}=b^{-1}\Pi_{\Phi}$, 
$v=b^{-1}u$, $\Pi_{v}=b^{-1}\Pi_{u}$ and $\bv=b^{-1}\bu$, 
$\Pi_{\bv}=b^{-1}\Pi_{\bu}$ will be used.

Introduce modes for $\vf$ by 
\begin{equation} \begin{split}
\vf(\si)=& q+i\sum_{n\neq 0}\frac{1}{n}
\bigl( a_n e^{-in\si} +\ba_n e^{in\si} \bigr) \\
\Pi_{\vf}(\si) =& 2P+\sum_{n\neq 0}
\bigl( a_n e^{-in\si} +\ba_n e^{in\si} \bigr) 
\end{split}\end{equation}
and modes for $v$, $\bv$ by
\begin{equation}
\begin{aligned}
v(\si)=& v+i\sum_{n\neq 0}\frac{1}{n}
v_ne^{in\si} \\
\Pi_{v}(\si)=&\Pi_{v,0}+\sum_{n\neq 0}
\Pi_{v,n}e^{in\si}
\end{aligned}\qquad
\begin{aligned}
\bv(\si)=& \bv-i\sum_{n\neq 0}\frac{1}{n}
\bv_ne^{-in\si} \\
\Pi_{\bv}(\si)=&\Pi_{\bv,0}+\sum_{n\neq 0}
\Pi_{\bv,n}e^{-in\si}
\end{aligned}
\end{equation}
The only nonvanishing commutators are
\begin{equation}\begin{aligned}
{[}P,q{]}&=-\fr{i}{2} \\
{[}\Pi_{v,0},v{]}&=-i \\
{[}\Pi_{\bv,0},\bv{]}&=-i
\end{aligned}\qquad
\begin{aligned}
{[}a_n,a_m{]}&=\fr{n}{2}\de_{n,-m} \\
{[}\ba_n,\ba_m{]}&=\fr{n}{2}\de_{n,-m} 
\end{aligned}\qquad
\begin{aligned}
{[}v_n,\Pi_{v,m}{]}&=n\de_{n,-m} \\
{[}\bv_n,\Pi_{\bv,m}{]}&=-n\de_{n,-m}
\end{aligned}
\end{equation}
Let $\CF$ be the Fock space generated from the Fock vacuum $\Om$ defined by
\begin{equation}\begin{aligned}
a_n\Om=0 \\
\ba_n\Om=0
\end{aligned}
\begin{aligned}\qquad
v_n \Om=0 \\
\Pi_{v,n}\Om=0
\end{aligned}
\begin{aligned}\qquad
\bv_n \Om=0 \\
\Pi_{\bv,n}\Om=0
\end{aligned}
\qquad \text{for }n>0
\end{equation}
The quantum Hamiltonian, normal ordered corresponding to the choice of 
Fock vacuum then reads
\begin{equation}\label{Ham}
\begin{aligned}
H= & H_{\sst F}+H_{\sst I} \\
H_{\sst F} =& 2(P^2+ibP)+
              \sum_{k=1}^{\infty}\bigl(2a_{-k}a_k+2\ba_{-k}\ba_k+
\Pi_{v,-k}v_k+v_{-k}\Pi_{v,k} -\Pi_{\bv,-k}\bv_k-\bv_{-k}\Pi_{\bv,k}
       \bigr)\\
H_{\sst I}=& 2\int_{0}^{2\pi}\frac{d\si}{2\pi}\Pi_v(\si)\Pi_{\bv}(\si)
       :e^{2b\vf}:
\end{aligned}\end{equation} 
The operator $H$ will initially be considered to act in the space $\CS$ of all
\begin{equation}
\Psi=\sum_{I}\CA_I\Om\;\Psi^I(q,v,\bv),
\end{equation}
where the $\CA_I$ are monomials in the oscillators labelled by suitable
multi-indices $I$, and the functions $\Psi^I(h)\in\CC^{\infty}_c(H_3^+)$ are
assumed to be nonzero only for finitely many $I$.

One has to note however the following problem:
There does not seem to be a simple
scalar product that makes the Hamiltonian symmetric.
Instead one has an indefinite, but nondegenerate hermitian form $(.,.)$
w.r.t. which the Hamiltonian is symmetric:
First define a hermitian form $(.,.)_{\CF}$ on $\CF$ by the following 
hermiticity relations for the oscillators:
\begin{equation}
a_n^{\dagger}=\ba_{-n}\qquad 
\Pi_{v,n}^{\dagger}=\Pi_{\bv,-n}\qquad v_{n}^{\dagger}=-\bv_{-n}
\end{equation}
The hermitian form thereby defined is diagonalized by forming the linear
combinations
\begin{equation}
\begin{aligned}
e^{\pm}_n=&\fr{i}{2}\bigl( v_n+\bv_n\mp(\Pi_{v,n}-\Pi_{\bv,n})\bigr) \\
f^{\pm}_n=&\fr{1}{2}\bigl( v_n-\bv_n\pm(\Pi_{v,n}+\Pi_{\bv,n})\bigr) 
\end{aligned}\qquad
b^{+}_n=a_n+\ba_n,\quad b^{-}_n=i(a_n-\ba_n),
\end{equation}
which diagonalize algebra and hermiticity relations:
\begin{equation}
\begin{aligned}
{[}e^{\pm}_n,e^{\pm}_m{]}=&\pm n\de_{n+m}\\
{[}f^{\pm}_n,f^{\pm}_m{]}=&\pm n\de_{n+m}\\
{[}b^{\pm}_n,b^{\pm}_m{]}=&\pm n\de_{n+m}
\end{aligned}\qquad\qquad
\begin{aligned}
(e^{\pm}_n)^{\dagger}=& e^{\pm}_{-n} \\
(f^{\pm}_n)^{\dagger}=& f^{\pm}_{-n} \\
(b^{\pm}_n)^{\dagger}=& b^{\pm}_{-n}.
\end{aligned}\end{equation}
It is easy to see that only the oscillators with 
superscript ($-$) generate elements $F$ of negative norm $(F,F)_{\CF}$.
The form $(.,.)_{\CF}$ is then diagonal on monomials $\CB_I$ in the 
oscillators $e^{\pm}_{-n}$, $f^{\pm}_{-n}$, $b^{\pm}_{-n}$, labelled by 
a suitable multi-index $I$:
\begin{equation}
(\CB_I\Om,\CB_J\Om)_{\CF}=\de_{I,J}\CN_I.
\end{equation}

The hermitian form $(.,.)$ is then defined on $\CS$ by
\begin{equation}
(\Psi_2,\Psi_1)=\sum_{I}\CN_I
\bigl(\Psi_2^I,\Psi_1^I\bigr)_{H_3^+},
\end{equation}
where $(.,.)_{H_3^+}$ is the $SL(2,\BC)$-invariant scalar product on $H_3^+$,
\begin{equation}
(\Psi_2^I,\Psi_1^I)_{H_3^+}=\int_{\BR}dq e^{-2bq} \int_{\BC}d^2 v 
\;\bigl(\Psi_2^I(q,v,\bv)\bigr)^{\ast}\Psi_1^I(q,v,\bv)
\end{equation}

Even though $(\Psi,\Psi)$ is clearly indefinite, there is a canonical norm
$\lVert . \rVert$ associated to it:
\begin{equation}
\lVert\Psi\rVert^2=\sum_{I}\;\lvert\CN_I
\rvert\,
\lVert\Psi^I(h)\rVert_{H_3^+}^2.
\end{equation}
Note that $\lVert\Psi\rVert^2\neq (\Psi,\Psi)$ due to taking the absolute
value of $(\CA_{I}\Om,\CA_{I}\Om)_{\CF^{}}$. 

One may then finally define the space
of states $\CV$ as the completion of $\CS$ w.r.t. $\lVert . \rVert$
 
\subsection{Current algebra}

In terms of the canonical variables one
may construct the currents
\begin{equation}\label{genL}
\begin{split}
J^{-}(\si)=& i\Pi_{\bv}(\si) \\
J^{0}(\si)=& +i:\bv(\si)\Pi_{\bv}(\si):+
               i\fr{b^{-1}}{2}\bigl(\Pi_{\vf}(\si)+\vf'(\si)\bigr) \\
J^{+}(\si)=& -i\Bigl(k\bv'(\si)+:(\bv(\si))^2\Pi_{\bv}(\si):+b^{-1}
               \bv(\si)\bigl(\Pi_{\vf}(\si)+\vf'(\si)\bigr)+
               \Pi_v(\si):e^{2b\vf(\si)}:\Bigr),
\end{split}
\end{equation}
as well as currents $\bJ^a(\si)$ obtained by hermitian conjugation.
The two crucial properties satisfied by these definitions are:\\
(1) {\it The currents have the following commutation relations with $H$:
\begin{equation}
[H,J^a(\si)]=-i\pa_{\si}^{}J^a(\si) \qquad [H,\bJ^a(\si)]=i\pa_{\si}^{}
\bJ^a(\si).
\end{equation}
From this it follows that the equations of motion of these observables
are simply}
\begin{equation}
\pa_{-}J^a(\si,\tau)=0\qquad \pa_{+}\bJ^a(\si,\tau)=0
\end{equation}
(2) {\it These currents generate a 
$\fsl_2\oplus \fsl_2$ current algebra with central charge $k$ related to $b$ 
by $b^{-2}=-(k+2)$: The modes $J_n^a$ of $J^a(\si)$, 
$a=-,0,+$ defined by $J^a(\si)=\sum_ne^{-in\si}J_n^a$ satisfy
\begin{equation}\label{algebra}
\begin{aligned} {[}J_n^0,J_m^0{]}& =\fr{k}{2}n\de_{n+m,0} \\ 
{[}J_n^0,J_m^{\pm}{]}& =\pm J_{n+m}^{\pm} \end{aligned} \qquad
{[}J_n^{+},J_m^{-}{]}=2J_{n+m}^0+kn\de_{n+m,0},  
\end{equation}
the modes $\bar{J}_n^a$ of $\bar{J}^a(\si)$ commute with the 
$J_n^a$ and 
satisfy the same algebra.} 

The verification of these assertions may be simplified by observing 
that the currents $J^a_{\sst F}$ obtained from the 
expressions \rf{genL} by dropping the terms containing $e^{2b\vf}$ 
are similiar to the standard free field
constuctions \cite{Wa}\cite{BF}\cite{FF}\cite{BO}
of $\fsl_2$ current algebras. 
These free field currents $J^a_{\sst F}$ may be deformed as
\begin{equation}\label{deform}\begin{split}
J^+_{{\mu}}(\si)=&J^+_{\sst F}(\si)+\mu \Pi_v(\si):e^{2b\vf(\si)}: \\
\bJ^+_{{\mu}}(\si)=&
\bJ^+_{\sst F}(\si)+ \mu \Pi_{\bv}(\si):e^{2b\vf(\si)}:,
\end{split}\end{equation}
without changing the algebra. The check that the modes of
$J^a_{{\mu}}(\si)$ commute with those of $\bJ^a_{{\mu}}(\si)$ 
essentially boils down to the fact that the deformation of  
$J^-_{\sst F}(\si)$ is the screening charge for the algebra 
generated by $\bJ^-_{\sst F}(\si)$ and vice versa.

One should note however an 
important difference between the free field representation that appears here
and the usual free field representation
of $\fsl_2$ current algebra: The definition of the Fock vacuum $\Om$ does
not involve the condition $\Pi_{v,0}\Om=0$, $\Pi_{\bv,0}\Om=0$, which
is usually imposed to get highest weight representations of the current
algebra. Not imposing these conditions is due to the fact that neither
$v$ nor $\Pi_{v,0}$ take values in compact sets so will get continuous
spectra upon quantization.
One will therefore have to deal with current algebra representations that
are neither of highest nor lowest weight type.

Associated with the current algebras one has two commuting Virasoro algebras
by the Sugawara construction: In terms of the canonical variables
the standard Sugawara energy-momentum tensors express as
\begin{equation}
T_{{\mu}}(\si)=+:\Pi_{\bv}\bv':+\fr{1}{4}:\bigl(\Pi_{\vf}+\vf'\bigr)^2:
-\fr{b}{2}e^{i\si}\pa_{\si}e^{-i\si}\bigl(\Pi_{\vf}+\vf'\bigr)
+\mu \Pi_{\bv}\Pi_{v}:e^{2b\vf}:
\end{equation}
and the corresponding expression for $\bT_{{\mu}}$. The operator 
\[ H_{{\mu}}=
\int_{0}^{2\pi}\frac{d\si}{2\pi}\bigl(T_{{\mu}}(\si)+\bT_{{\mu}}(\si)
\bigr) \]
satisfies 
\begin{equation}
[H_{{\mu}},J^a_{{\mu}}(\si)]
=-i\pa_{\si}^{}J^a_{{\mu}}(\si) \qquad 
[H_{{\mu}},\bJ^a_{{\mu}}(\si)]=i\pa_{\si}^{}\bJ^a_{{\mu}}(\si)
\end{equation}
for any value of $\mu$. By taking the value $\mu=1$ one recovers the
canonical Hamiltonian of the $H_3^+$-WZNW model.

\subsection{The Spectrum}

The decompostion of the space of states into {\it irreducible} representations 
of the current algebra $\fsl_2\oplus\fsl_2$ was found in \cite{Ga} by 
explicitely evaluating the path integral on the torus. The result was then
shown to agree with the definition of the partition function via
\begin{equation}
\CZ(q,u)=|q|^{-\frac{c_k}{12}}\Tr_{\CH} q^{L_0}\bq^{\bL_0} 
e^{2\pi i(uJ_0^0-\bu\bJ_0^0)}
\end{equation}
if the trace is performed over the space $\CH$ spanned by $L^2(H_3^+)$
and the descendants obtained by repeated action of the 
$J_n^a$, $\bJ_n^a$  with $n<0$. If $\fn_-$ is the
Lie subalgebra spanned by the $J_n^a$, $\bJ_n^a$  with $n<0$ and 
Sym($\fn_-$) the corresponding symmetric algebra one may
write the definition of $\CH$ formally as
\begin{equation}
\CH=\text{Sym($\fn_-$)}\ot L^2(H_3^+)
\end{equation}

In order to introduce an action of the current algebra in $\CH$
note that any representation $V$ of a finite dimensional Lie algebra $\fg$ 
(generators $J^a$) can be 
extended to a representation $\CP(V,\fg,k)$ of the affine algebra $\hfg$ 
(generators $J^a_n$) corresponding to $\fg$
by demanding $J_n^a V=0$, $n>0$ and by 
extending $V$ by the linear span of expressions of the form 
\[ J^{a_1}_{-n_1}\ldots J^{a_k}_{-n_k}v \quad\mbox{for any}\quad v\in V. \]
These kind of current algebra representations have been named 
prolongation modules in \cite{LZ}. To write the definition of $\CP(V,\fg,k)$ 
more precisely, introduce the Lie subalgebra $\fb_+$ spanned by the $J_n^a$, 
$\bJ_n^a$  with $n\geq 0$. One then has
\begin{equation}
\CP(V,\fg,k)=\CU(\hfg)\ot_{\CU(\fb_+)}V
\end{equation}
  
In the present case one may choose $V$ to be a suitable 
dense subspace $\CS(H_3^+)$
of $L^2(H_3^+)$, which is invariant under the action of the zero mode
subalgebra $\fg=\fsl_2\oplus\fsl_2$ by the differential operators
\begin{equation}\label{diffrep}
\begin{aligned}
J^+= &-e^{2\vf}\frac{\pa}{\pa v}
-\bv^2\frac{\pa}{\pa \bv}-\bv \frac{\pa}{\pa \vf}
      \\
\bJ^+= &-e^{2\vf}\frac{\pa}{\pa \bv}
-v^2\frac{\pa}{\pa v}-v \frac{\pa}{\pa \vf}
        \end{aligned}\qquad 
\begin{aligned}
J^0= &\bv\frac{\pa}{\pa \bv}+\frac{1}{2}\frac{\pa}{\pa \vf} \\
\bJ^0= &v\frac{\pa}{\pa v}+\frac{1}{2}\frac{\pa}{\pa \vf}
\end{aligned}\qquad 
\begin{aligned}
J^-= &\frac{\pa}{\pa \bv} \\
\bJ^-= &\frac{\pa}{\pa v} .
\end{aligned} 
\end{equation} 
The corresponding prolongation module
\begin{equation}
\hat{\CS}(H_3^+)\equiv \CP\bigl(\CS(H_3^+),\fsl_2\oplus\fsl_2,k\bigr)
\end{equation}
is then a dense subspace of $\CH$ on which the current algebra 
$\fg=\hfsl_2\oplus\hfsl_2$ is represented.

The decomposition of $L_2(H_3^+)$ into irreducible
representations \cite{GGV} (see also discussion in \cite{T1}) reads
\begin{equation}\label{abstdecom} \CH\equiv L^2(H_3^+,dh)=
\int_{\rho>0}^{\oplus} d\rho \rho^2\;\,\CH_{-\frac{1}{2}+i\rho} ,
\end{equation}
where $ \CH_j$ is a representation of the princial series of $SL(2,\BC)$.
This decomposition then induces a corresponding decomposition of
$ \hat{\CS}(H_3^+)$ into current-algebra representations 
\begin{equation}
\CP_j\equiv \CP\bigl(\CH_j,\fsl_2\oplus\fsl_2,k\bigr), 
\end{equation}
which are irreducible by the results in \cite{T2}. It is important to note that
although one has chiral factorization on the level of the algebra, one has no
factorization of the representation $\CP_j$ in the form $\CR_j\ot\bar{\CR}_j$,
where $\CR_j$ (resp. $\bar{\CR}_j$) are irreducible representations of the 
current algebras generated by $J_n^a$ (resp. $\bar{J}_n^a$).  

By the Sugawara construction one then also gets an
action of two commuting Virasoro algebras (generators $L_n$, $\bL_n$) on 
$ \hat{\CS}(H_3^+)$ resp. $\CP_j$. States $\Psi\in \CS_3^+$ satisfy 
\begin{equation}
L_0 \Psi =-t^{-1}Q\Psi, \quad \bL_0 \Psi =-t^{-1}Q\Psi, 
\quad\text{and }L_n \Psi =0, \quad
\bL_n\Psi=0\quad\text{for each } n >0,
\end{equation}
where $Q$ is the Laplacian on \hp, $t=-(k+2)$.
They will also be called lowest level or {\it primary} states. 
Primary states $\Psi_j\in\CP_j$ satisfy $L_0 \Psi =h_j\Psi$ with
$h_j=t^{-1}j(j+1)$. 

Let me finish this section by noting that the representations $\CP_j$ 
are equivalent to a representation $\CF_j$
of the current algebra $\hfg$ in the space
$\CF\ot\CS(\BC)$ by means of the modes $I^a_n$ of
\begin{equation}\label{genFF}
\begin{split}
I^{-}(\si)=& i\Pi_{\bv}(\si) \\
I^{0}(\si)=& +i:\bv(\si)\Pi_{\bv}(\si):+
               i\fr{b^{-1}}{2}\bigl(\Pi_{\vf}(\si)+\vf'(\si)\bigr) \\
I^{+}(\si)=& -i\Bigl(k\bv'(\si)+:(\bv(\si))^2\Pi_{\bv}(\si):+b^{-1}
               \bv(\si)\bigl(\Pi_{\vf}(\si)+\vf'(\si)\bigr)\Bigr),
\end{split}\end{equation}
as well as their ``antiholomorphic'' counterparts $\bI^a_n$ defined 
analogously.
The generators $v_n$, $\bv_n$, $\Pi_{v,n}$, $\Pi_{\bv,n}$, $a_n$, $\ba_n$
as well as the definition of the Fock-space $\CF$ are as in section 2.1.
The zero mode generator $P$ acts by mutliplication with $ibj$. 

\section{The bootstrap}

The construction and calculation of correlation functions is 
not easy to achieve directly in the framework of canonical quantization
since no explicit construction of primary states or fields is known.

Instead it has in the case of RCFT turned out to be extremely useful
to exploit as much as possible the Ward identities 
from the current algebra symmetries by following a strategy 
similar to that introduced in \cite{BPZ}. The aim of the 
present section will be to generalize the usual formalism to the  
case where one is no longer dealing with lowest- or highest 
weight representations.

\subsection{Primary and secondary fields} 
   
Primary fields $\Phi[f|z)$ can be associated to each vector in the zero mode
representation, here sufficiently differentiable
functions $f$ on $H_3^+$. The vector $f$ to which 
$\Phi[f|z)$ corresponds is recovered by the usual prescription for 
operator-state correpondence:
\begin{equation}\label{op-state}
 \lim_{z\ra 0}\Phi[f|z)=f
\end{equation}
They transform under the current algebra in a particularly simple way:
\begin{equation}\label{prim1}
[J_n^a,\Phi[f|z)]=z^n\Phi[J_0^a f|z)\qquad
[\bJ_n^a,\Phi[f|\si)]=\bz^n\Phi[\bJ_0^af|z),
\end{equation}
where the action of $J_0^a$, $\bJ_0^a$ on $f$ is by \rf{diffrep}. 

A convenient plane-wave normalizable basis for $L^2(H_3^+)$ was in 
\cite{GGV}, see also \cite{T1}, shown to be given by the functions
\begin{equation}
\Psi(j;x|h)=\frac{2j+1}{\pi}\left((1,x)\cdot h\cdot \binom{1}{\bx}\right)^{2j}
\end{equation}
The corresponding primary fields $\Phi[ \Psi(j;x|.)|z)$ will be
denoted $\Phi^j(x|z)$. The transformation law \rf{prim1} may then be 
reformulated as the following OPE
\begin{equation}\label{prim2}
  J^a(z)\Phi^j(x|w)=\frac{1}{z-w}      \CD_j^a \Phi^j(x|w),\qquad 
\bJ^a(\bz)\Phi^j(x|w)=\frac{1}{\bz-\bw}\bar{\CD}_j^a \Phi^j(x|w)
\end{equation}
where the differential operators $\CD_j^a$
representing $\fsl_2$ are
\begin{equation}
\CD^+_j=-\bx^2\pa_{\bx}+2j\bx \qquad \CD^0_j=-\bx\pa_{\bx}-j 
\qquad \CD^-_j=\pa_{\bx},
\end{equation}
the $\bar{\CD}_j^a$ their complex conjugates. 

The representations with spin $j$ and $-j-1$ are equivalent for $j\notin\BZ$.
This implies that the operators $\Phi^{j}(x|z)$ and $\Phi^{-j-1}(x|z)$ must be 
related by a relation of the form 
\begin{equation}
\Phi^{-j-1}(x|z)=-R(j)\frac{2j+1}{\pi}\int_{\BC}d^2x'\; |x-x'|^{-4j-4}
\Phi^{j}(x'|z),
\end{equation}
thereby defining a ``reflection'' amplitude $R(j)$ not restricted by current
algebra symmetry.

Descendant (secondary) fields will be defined for each monomial 
$\CJ_{I_L}\bar{\CJ}_{I_R}$ where
$\CJ_I\equiv J_{-n_1}^{a_1}\ldots J_{-n_k}^{a_k}$ for multi-index
$I=(n_1,\ldots,n_i,\ldots,n_k;a_1,\ldots,a_i,\ldots,a_k)$, $n_i<0$, 
and correspondingly for $\bar{\CJ}_{I}$:
\begin{equation}\begin{split}
\bigl[\CJ_{I_L}\bar{\CJ}_{I_R}\Phi\bigl](x|z)\equiv 
\prod_i \frac{1}{(n_i-1)!}: & \bigl((\pa_z)^{n_1-1}J^{a_1}(z)\bigr)
\ldots \bigl((\pa_z)^{n_k-1}J^{a_k}(z)\bigr)\\
 & \bigl((\pa_{\bz})^{\bn_1-1}J^{\ba_1}(\bz)\bigr)
\ldots \bigl((\pa_{\bz})^{\bn_{\bk}-1}J^{\ba_{\bk}}(\bz)\bigr)\Phi^j(x|z):
\end{split}\label{descdef}\end{equation}

\subsection{Correlation functions}

The assumption of invariance of correlation functions under the 
symmetries generated by $J_0^a$, $\bJ_0^a$, $L_n$, $\bL_n$, $n=-1,0,1$
determines two and three-point functions up to certain functions of the 
$j_i$:
\begin{equation}\begin{split}
<\Phi^{j_2}(x_2|z_2)&\Phi^{j_1}(x_1|z_1)>
\begin{aligned}[t]
=& N(j_1)|z_1-z_2|^{4h_1}\de^{(2)}(x_1-x_2)\delta(j_1,-j_2-1)\\
+& B(j_1)|z_1-z_2|^{4h_1}|x_1-x_2|^{4j_1}\delta(j_1,j_2) 
\end{aligned} \\
<\Phi^{j_3}(x_3|z_3)&\Phi^{j_2}(x_2|z_2)
\Phi^{j_1}(x_1|z_1)>= \\
& |x_1-x_2|^{2(j_1+j_2-j_3)}|x_1-x_3|^{2(j_1+j_3-j_2)}
|x_2-x_3|^{2(j_2+j_3-j_1)}\ti \\
 & |z_1-z_2|^{2(h_3-h_1-h_2)}|z_1-z_3|^{2(h_2-h_1-h_3)}
|z_2-z_3|^{2(h_1-h_2-h_3)}C(j_1,j_2,j_3), 
\end{split}\end{equation}
where $h_i=h(j_i)$, $i=1,2,3$. The two terms in the two point function again 
arise due to the equivalence of representations with spin $j $ and $-j-1$.
I will assume the operators to be normalized by 
\begin{equation}
N(j)\equiv 1 \qquad \text{such that}\qquad R(j)=B(j)
\end{equation} 
Furthermore I will assume the $C(j_1,j_2,j_3)$ to be symmetric in its variables
as is necessary for the primary fields to be mutually local. 

Correlation functions of descendant fields
may as usually be reduced to those of primary fields by using \rf{descdef}
and the OPE \rf{prim2}.

In order to use the current algebra symmetries to get information on 
n-point functions for $n>3$ one postulates operator product expansions
of the form 
\begin{equation}\label{OPE}\begin{split}
\Phi^{j_2}(x_2|z_2)\Phi^{j_1}(x_1|z_1)=
\int_{\CS} d\mu(j) & \; |z_2-z_1|^{2(h_j-h_2-h_1)}\\
 & \ti\sum_{n,\bn=0}^{\infty}(z_2-z_1)^n(z_2-z_1)^{\bn} 
\CO_{n\bn}^{j}\bigl[\,{}^{j_2}_{x_2}{}^{j_1}_{x_1}\bigr](z_1).
\end{split}\end{equation}
The measure $d\mu(j)$ was introduced to simplify 
notation as one will in general have to sum over
discrete as well as continuous sets of $j$. 

Requiring that both sides of \rf{OPE} transform the same way under the 
current algebra allows to express the operators $\CO_{n\bn}^{j}$ 
as linear combinations of $\Phi^{j}(x|z_1)$ and its descendants \cite{T2}:
\begin{equation}
\CO_{n\bn}^{j}\bigl[{}^{j_2}_{x_2}{}^{j_1}_{x_1}\bigr](z_1)
= \sum_{I_L\in\mathbf{I}_n}\sum_{I_R\in\mathbf{I}_{\bn}}\int_{\BC}d^2x\;
\CC_{I_LI_R}\!\left(\,{}^{j}_{x}{}^{j_2}_{x_2}{}^{j_1}_{x_1}\right)
\bigl[\CJ_{I_L}\bar{\CJ}_{I_R}\Phi\bigr](x|z_1),
\end{equation}
where the set $\mathbf{I}_n$ contains all multi-indices 
$I=(n_1,\ldots,n_k;a_1,\ldots,a_k)$ such that $n=\sum_i n_i$ and
the coefficients $\CC_{I_LI_R}$ are uniquely defined in terms
of $\CC_{0,0}$, which reads
\begin{equation}\begin{split}
\CC_{0,0}\left(
\begin{smallmatrix} j & j_2 & j_1\\ x & x_2 & x_1 \end{smallmatrix}
\right)=&
D(j;j_2,j_1)C\left(
\begin{smallmatrix} -j-1 & j_2 & j_1\\ x & x_2 & x_1 \end{smallmatrix}
\right) \\
\equiv &  D(j;j_2,j_1) |x_1-x_2|^{2(j_1+j_2+j+1)}|x_1-x_3|^{2(j_1-j_2-j-1)}
|x_2-x_3|^{2(j_2-j_1-j-1)}
\end{split}\end{equation}
The only remaining freedom is given by the structure constants 
$D(j;j_2,j_1)$. In order to define them uniquely, one has to assume that only
one of the two linearly dependent operators $\Phi^{j}(x|z_1)$ and 
$\Phi^{-j-1}(x|z_1)$ appears in \rf{OPE}. One might i.e. take the integration
region $\CS$ in \rf{OPE} as subset of $\bigl\{j\in\BC; \,\arg(2j+1)\in\bigl(
-\fr{\pi}{2},\fr{\pi}{2}\bigr]\bigr\}$. 
By using \rf{OPE} in a three point function one then finds that 
\begin{equation}
D(j;j_2,j_1)=C(-j-1,j_2,j_1).
\end{equation}
 
Four point functions may then be expanded in terms of three point functions
by i.e. using the OPE \rf{OPE} of operators $\Phi^{j_2}(x_2|z_2)$ 
and $\Phi^{j_1}(x_1|z_1)$. One 
arrives at a representation of the four point function in the form
\begin{eqnarray}\lefteqn{
<\Phi^{j_4}(x_4|z_4) \ldots \Phi^{j_1}(x_1|z_1)>= }
\nonumber \\
&= \dst \int_{\CS_{s}}d\mu(j_{21})\;\;\; C(j_4,j_3,j_{21})D(j_{21};j_2,j_1) &
\left|\CF_{s,j_{21}}  \Big[
{}^{j_4}_{x_4}{}^{j_3}_{x_3}{}^{j_2}_{x_2}{}^{j_1}_{x_1}\Big]\!
\dst \left(z_4,\ldots z_1\right)\right|^2
\label{ans} 
\end{eqnarray}
This representation splits the information involved in the
definition of the four point function into a piece determined 
directly by the current algebra symmetries (the conformal blocks 
$\CF_{s,j_{21}}$, the subscript s refers to the ``s-channel'')
and two pieces of information that one should expect to be determined in
terms of the conformal blocks only rather indirectly: The structure 
constants $C(j_3,j_2,j_{1})$ and the set $\CS_{s}$ of intermediate 
representations. The latter is of course equivalent to knowledge of the
fusion rules, i.e. the set of representations appearing in operator
product expansions.

\subsection{Crossing symmetry}

An alternative representation of the four
point function is obtained by using the OPE \rf{OPE} of operators 
$\Phi^{j_3}(x_3|z_3)$ and $\Phi^{j_2}(x_2|z_2)$ to get an expansion in terms
of ``t-channel'' conformal blocks:
\begin{eqnarray}\lefteqn{
<\Phi^{j_4}(x_4|z_4) \ldots \Phi^{j_1}(x_1|z_1)>= }
\nonumber \\
&= \dst \int_{\CS_{t}}d\mu(j_{32})\;\;\; C(j_4,j_1,j_{32})D(j_{32};j_3,j_2) &
\left|\CF_{t,j_{32}}  \Big[
{}^{j_4}_{x_4}{}^{j_3}_{x_3}{}^{j_2}_{x_2}{}^{j_1}_{x_1}\Big]\!
\dst \left(z_4,\ldots z_1\right)\right|^2
\label{ant} 
\end{eqnarray}

A fundamental physical requirement, which is equivalent to mutual locality
of primary fields, is that the expansions \rf{ans} and \rf{ant} into 
s-channel and t-channel conformal blocks produce the same correlation 
functions (crossing symmetry).
One may hope to infer from the equality of the two decompositions \rf{ans},
\rf{ant} the existence of fusion relations by an argument similar to that
given for RCFT in \cite{MS}:
\begin{equation}\label{deffus}
\CF_{s,j_{21}} \Big[
{}^{j_4}_{x_4}{}^{j_3}_{x_3}{}^{j_2}_{x_2}{}^{j_1}_{x_1}\Big]\!
(z_4,\ldots,z_1)=\dst
\int d\mu(j_{32})\;\; F_{j_{21}j_{32}}\!\!\left[{}^{j_3j_2}_{j_4j_1}\right]
\CF_{t,j_{32}}  \Big[
{}^{j_4}_{x_4}{}^{j_2}_{x_2}{}^{j_3}_{x_3}{}^{j_1}_{x_1}\Big]\!
(z_4,\ldots,z_1)
\end{equation}
Indeed, in order to make up an argument of the type given in \cite{MS} one
only needs {\it existence} of an extension of the set of conformal blocks
$\CF_{t,j_{32}}$ to a basis, with respect to which the $\CF_{s,j_{21}}$ can be
expanded. Let me note that existence of fusion relations in the
mini-superspace limit was shown in \cite{T1}.

Given fusion relations \rf{deffus} the requirement of crossing
symmetry translates 
itself into a system of equations for the structure constants:
\begin{equation}\label{cross-gen}
\begin{split}
\dst\int_{\CS_s} d\mu(j_{21}) 
\;F_{j_{21}^{}j_{32}^{}}\!\!\left[{}^{j_3j_2}_{j_4j_1}\right]
\bar{F}_{j_{21}^{}j_{32}'}\!\!\left[{}^{j_3j_2}_{j_4j_1}\right]
\;  & C(j_4,j_3,j_{21})D(j_{21};j_2,j_1) \\
= \delta(j_{32}^{},j_{32}')\; & C(j_4,j_{32},j_1)
C(j_{32};j_3,j_2).
\end{split}\end{equation}
Viewing the fusion transformations as being given by the conformal blocks, 
therefore indirectly from the current algebra symmetries, one should read
\rf{cross-gen} as possible starting point for the determination of the
structure constants.

\section{Structure constants}

The aim of the present section will be to derive an explicit expression for the
structure constants $C(j_3,j_2,j_1)$ that appear in the expansion 
(\ref{ans}) for the four point function of four arbitrary primary fields.

\subsection{Degenerate fields}

The representations $\CP_j$ are irreducible for generic $j$. They become 
degenerate if and only if $j$ equals any of the $j_{r,s}$, where
\begin{equation}
2j_{r,s}+1=r-st \text{ where }\begin{cases}
\text{either }&r\geq 1, s\geq 0\\
\text{or }& r< -1, s< 0
\end{cases}  
\end{equation} 
In a formal sense the degeneracy may be seen to arise due to the
existence of null vectors. However, these are not found among the normalizable
vectors of the representations $\CP_j$
but are distributional objects instead, cf. 
\cite{T2}.

Correspondingly there 
exist fields that satisfy additional differential equations
\cite{T2}. These fields 
will be called degenerate fields in the following and
denoted $\Phi_{r,s}$.
I will need only the following two simple examples of degenerate fields, 
corresponding to 
$j=j_{2,1}=1/2$ and $j_{1,2}=-t/2$ respectively. 

The degenerate field $\Phi_{2,1}(x|z)$ satisfies
\begin{equation}\label{deg21}
\pa_x^2\Phi_{2,1}(x|z)=0 \qquad\pa_{\bx}^2\Phi_{2,1}(x|z)=0.
\end{equation} 
It transforms in the 
finite dimensional spin 1/2 representation of $SL(2,\BC)$, and is therefore
identified with the quantum analogue of the fundamental field $h$.

In the other case $\Phi_{1,2}(x|z)$
the differential equation expressing degeneracy reads
\begin{equation}\label{deg12}
: \Bigl( J^+(x|z)\pa_x^2-2(1+t)J^0(x|z)\pa_x-t(1+t)J^-(x|z) \Bigr)
\Phi_{1,2}(x|z) :\;\; =\; 0,
\end{equation}
where $J^a(x|z)=e^{xJ^-_0}J^a(z)e^{-xJ^-_0}$.

\subsection{The method, assumptions}

The method to be used will consist of
considering four point functions in which one of the operators
is a degenerate field $\Phi_{r,s}$, the others generic.
The main assumption will be that these four point functions 
are crossing symmetric.
More precisely the assumption is that the conditions \rf{cross-gen}
for crossing symmetry with four generic fields are
compatible with those considered here.

The considered four point functions will satisfy degeneracy equations 
in addition to the KZ equations. The set of conformal blocks that solves 
both equations will be finite. Moreover one has 
fusion and braiding relations relating different bases for the 
conformal blocks corresponding to the different possible 
factorization patterns. 

The requirement of crossing symmetry
thereby takes the form
\begin{equation}\label{cross-spe}
\begin{array}{ccc}
\multicolumn{2}{l}{
\dst\sum_{s\in\mathbf{F}_{r,s}} \;\,C(j_4,j_3,j_1+s)D(j_1+s;j_2,j_1)
\;F_{st}\!\!\left[{}^{j_3j_2}_{j_4j_1}\right]
F_{st'}\!\!\left[{}^{j_3j_2}_{j_4j_1}\right]} 
& \qquad\qquad  \\[2ex]
\qquad\qquad  & 
\multicolumn{2}{r}{= \;\delta_{t,t'}\;C(j_4,j_1,j_3+t)
D(j_3+t;j_3,j_2).}
\end{array}\end{equation}
For both cases $\Phi_{2,1}$ and $\Phi_{1,2}$ it will be possible to
calculate conformal blocks and fusion matrices explicitely.
Given matrices
$ F_{st}$, the equations (\ref{cross-spe}) are finite difference
equations for the unknown $C(j_3,j_2,j_1)$.
It will be shown that a solution to both the equations from 
$j_{2}=1/2$ and $j_2=-t/2$ exists and is unique when $t$ is irrational. 

\subsection{Differential equations for the conformal blocks}

It was shown in \cite{T2} that the conformal blocks of any collection
of primary fields satisfy a generalization of the KZ equation previously
introduced in \cite{FZ}. In terms of the cross-ratios $x,z$ the KZ
equation takes the form
\begin{equation}\label{kzred}\begin{split}
tz(z-1)\pa_z \CF =& \CD_x^{(2)}\CF, \qquad\text{where} \\
\CD_x^{(2)} =& x(x-1)(x-z)\pa_x^2 \\
             & -((\D-1)(x^2-2zx+z)+2j_1x(z-1)+2j_2x(x-1)+2j_3z(x-1))\pa_x \\
             & +2j_2\D(x-z)+2j_1j_2(z-1)+2j_2j_3z
\end{split}\end{equation}

Conformal blocks for correlation functions involving degenerate fields 
satisfy additional differential equations:
 
In the case of the (2,1) degenerate field the decoupling equation reads 
simply $\pa_x^2\CF=0$. The two linearly independent solutions corresponding 
to s-channel conformal blocks with $j_{21}=j_1\pm 1/2$ are denoted
$\CF_{s,\pm}^{\sst 2,1}$. For the t-channel one has
$j_{32}=j_3\pm 1/2$ with notation $\CF_{t,\pm}^{\sst 2,1}$ respectively.
Their explicit expressions as well as the fusion matrices defined by
\begin{equation}\label{funot}
\CF_{s,\si}^{\sst 2,1}(x,z)=\sum_{\tau=\pm}F_{\si\tau}^{\sst 2,1} 
\CF_{t,\tau}^{\sst 2,1}(x,z)
\end{equation}
are given in the appendix.

The conformal blocks of a four point function 
involving the (1,2) degenerate field satsisfy a third order 
differential equation $\CD_x^{(3)}\CF=0$, where
\begin{equation}\label{12eqn}
\begin{split}
\CD_x^{(3)}=& x(x-1)(x-z)\pa_x^3 \\
            & -((\D-2)(x^2-2zx+z)+2j_1x(z-1)-2(1+t)x(x-1)+2j_3z(x-1))\pa_x^2 \\
            & -(2(1+t)(j_1(z-1)+j_3z-(\D-1)(z-x))-t(1+t)(x+z+1))\pa_x\\
            & -t(1+t)\D,
\end{split}\end{equation}
and $\D=j_1+j_2+j_3-j_4$. There are now three independent solutions 
in the s-channel case corresponding to $j_{21}=j_1\mp t/2$ as well as 
$j_{21}=-j_1-1+t/2$. They will be denoted $\CF_{s\pm}^{\sst 1,2}$ and 
$\CF_{s\ti}^{\sst 1,2}$ respectively. The fusion matrices 
$F_{\si\tau}^{\sst 1,2}$
are defined similarly as in \rf{funot}, with obvious modifications. 
Again the explicit expressions for conformal blocks and fusion matrices 
are to be found in the appendix.
 
\subsection{Liouville case}
It will turn out that parts of the analysis are closely related to the
corresponding analysis for Liouville theory \cite{TL}. This will not only 
allow to simplify some calculations but also shed some interesting light
on the relationship of these theories. 

The relation between the central charges is given by
\[ t=-b^{-2} \quad
\mbox{when the Liouville central charge is}\quad c_L=1+6Q^2, \quad 
Q=b+b^{-1} \]
whereas the relations between 
Liouville-momentum $\al$ resp. conformal dimension $h^{\sst L}_{\al}$
and the WZNW-spin $j$ are
\[ \al\equiv -bj; \qquad h^{\sst L}_{\al}\equiv \al(Q-\al) \]
The degenerate fields $\Phi^L_{2,1}$ and $\Phi^L_{1,2}$ have 
$\al_{2,1}=-b/2$ and
$\al_{1,2}=-b^{-1}/2$ respectively. The 
decoupling equation for the four-point function in case (2,1) reads
\begin{equation}\label{Lioudec}
\left(b^{-2}\pa_z^2+\frac{2z-1}{z(1-z)}\pa_z+\frac{h^{\sst L}_{\al_3}}{(1-z)^2}
+\frac{h^{\sst L}_{\al_1}}{z^2}+\frac{\kappa_L}{z(1-z)}
\right)\CF(z)=0, \end{equation}
where $\kappa_L=h^{\sst L}_{\al_1}+h^{\sst L}_{\al_2}+h^{\sst L}_{\al_3}-
h^{\sst L}_{\al_4}$.
For the case (1,2) one obtaines the corresponding equation by $b\ra b^{-1}$.
The notation for solutions, fusion matrices, structure constants etc. 
differs from that 
introduced previously just by adding a superscript L, i.e.
$\CF\ra\CF^{\sst L}$, $F\ra F^{\sst L}$. 

\subsection{Crossing symmetric combinations of conformal blocks}

It will be useful to reconsider the Liouville case 
along the lines of \cite{TL} first since it can be 
used to facilitate the analysis of the other cases.
\subsubsection{Liouville case revisited} 
The off-diagonal ($\ep\neq\ep'$) part of \rf{cross-spe} reads 
\begin{equation}\label{croff}
 \frac{E_{s+}^{{\sst L2,1}}}{E_{s-}^{{\sst L2,1}}}=
-\frac{F^{{\sst L2,1}}_{-+}F^{{\sst L2,1}}_{--}}{F^{{\sst L2,1}}_{+-}
F^{{\sst L2,1}}_{++}}, \qquad\mbox{where}\quad E_{s\si}^{{\sst L2,1}}\equiv
C^{\sst L}\left(\al_4,\al_3,\al_1-\si \fr{b}{2}\right)
C^{\sst L}\left(\al_1-\si\fr{b}{2} ;-\fr{b}{2} ,\al_1\right)
\end{equation}
where the right hand side is
explicitly given by ($\ga(y)=\frac{\Ga(y)}{\Ga(1-y)}$)
\[  \frac{F^{{\sst L2,1}}_{-+}F^{{\sst L2,1}}_{--}}{F^{{\sst L2,1}}_{+-}
F^{{\sst L2,1}}_{++}}=
-\frac{\ga\left(b(\al_1+\al_3-\al_4-\frac{b}{2})\right)
\ga\left(b(\al_1+\al_4-\al_3-\frac{b}{2})\right)
\ga\left(b(\al_1+\al_3+\al_4-\frac{3b}{2})-1\right)}
{\ga\bigl(b(2\al_1-b)\bigr)\ga\left(b(2\al_1-b)-1\right)
\ga\left(b(\al_3+\al_4-\al_1-\frac{b}{2})\right)}. \]
In order to determine $C_{\si}^{{\sst L2,1}}(\al_1)\equiv
C^{\sst L}\left(\al_1-\si\frac{b}{2} ;-\frac{b}{2},\al_1\right)$ 
one may consider the special case $\al_1=\al_4=\al$, 
$\al_3=-b/2$. Using that $C^{\sst L}(\al;\al_3,\al_{21})=
C^{\sst L}(Q-\al,\al_3,\al_{21})$ one now gets an equation that
involves $C_{\ep}^{{\sst L2,1}}(\al_1)$ only.
It is clear that the crossing symmetry relations \rf{cross-spe} can not 
determine the normalization 
of operators. The resulting freedom is fixed by imposing the normalization 
condition $C_+^{{\sst L2,1}}(\al)=1$.
Given that normalization, equations \rf{croff} determine 
$C_-^{{\sst L2,1}}(\al)$ to be
\[ C_-^{{\sst L2,1}}(\al)=\nu_{\sst L}(b)
\frac{\ga\bigl(b(2\al-b-b^{-1})\bigr)}{\ga(2b\al)}, \]
where $\mu_{\sst L}(b)$ represents the only leftover freedom, which 
corresponds to the cosmological constant.
Inserting this into \rf{croff} yields the functional relation for 
$C^{\sst L}(\al_3,\al_2,\al_1)$
derived in \cite{TL}:
\[  \frac{C^{\sst L}(\al_3,\al_2,\al_1+b)}{C^{\sst L}(\al_3,\al_2,\al_1)}=
\frac{\bigl(\mu{\sst L}(b)\bigr)^{-1}\ga(b(2\al_1+b))\ga(2b\al_1)
\ga(b(\al_3+\al_4-\al_1-b))}
{\ga(b(\al_1+\al_3-\al_4))\ga(b(\al_1+\al_4-\al_3))
\ga(b(\al_1+\al_3+\al_4-b)-1)}. \]
A second functional equation is of course obtained by $b\ra b^{-1}$
and $\mu_{\sst L}(b)\ra \tilde{\mu}_{\sst L}(b^{-1})$
These two functional equations are solved by an expression
of the form 
\[ C^{\sst L}(\al_1,\al_2,\al_3)=
\frac{(\mu_{\sst L}(b))^{-b^{-1}(\al_1+\al_2+\al_3)}\;
\up(2\al_1)\up(2\al_2)\up(2\al_3)}{\up(\al_1+\al_2+\al_3-b-b^{-1})
\up(\al_1+\al_2-\al_3)\up(\al_1+\al_3-\al_2)\up(\al_2+\al_3-\al_1)}, \]
if $\nu_{\sst L}$ and $\tilde{\nu}_{\sst L}$ are related by 
$(\tilde{\nu}_{\sst L}(b^{-1}))^b=(\nu_{\sst L}(b))^{b^{-1}}$ and
$\up(x)$ satisfies the functional relations 
$\up(x+b)=\ga(bx)b^{1-2bx}\up(x)$ and $b\ra b^{-1}$.
Such a function was introduced in \cite{DO} and \cite{ZZ}. 
In the latter reference the
function $\up(x)$ was defined by 
\[ \log\up(x)\equiv \int_{0}^{\infty}\frac{du}{u}
\left[\left(\frac{Q}{2}-x\right)^2e^{-u}
-\frac{\sinh^2\left(\frac{Q}{2}-x\right)\frac{u}{2}}{\sinh\frac{bu}{2}
\sinh\frac{b^{-1}u}{2}}
\right].\]
Taking into account the requirement of symmetry of 
$C^{\sst L}(\al_1,\al_2,\al_3)$ it was shown in \cite{TL} 
that for irrational values of $b$ the solution is unique up to a possibly 
$b$-dependent factor.

Let me summarize for future reference the properties of $\up(x)$ that will 
needed:
\begin{enumerate}
\item Symmetries $\up(Q-x)=\up(x)$, $\up_b(x)=\up_{b^{-1}}(x)$.
\item Functional relations $\up(x+b)=\ga(bx)b^{1-2bx}\up(x)$ and $b\ra b^{-1}$.
\item Poles at $x=x_{m,n}=-mb^{-1}-nb$ and $Q-x_{m,n}$ for 
$m,n\in\BZ^{\geq 0}$.
\end{enumerate}

\subsubsection{Case (2,1)}

The analysis is completely analogous to the Liouville case, with few changes:
\[  \frac{F^{\sst 2,1}_{-+}F^{\sst 2,1}_{--}}{F^{\sst 2,1}_{+-}
F^{\sst 2,1}_{++}}=-\frac{\ga(b(\al_1+\al_3-\al_4-b/2))
\ga(b(\al_1-\al_3+\al_4-b/2))
\ga(1-b(\al_3+\al_4-\al_1-b/2))}{\ga^2(b(2\al_1-b))
\ga(1-b(\al_1+\al_3+\al_4-3b/2))}. \]
Normalization condition $C^{\sst 2,1}_+(\al)=1$ and consideration of 
the special case $\al_3=-b/2$, $\al_1=\al_4$ now leads to 
\[ C^{\sst 2,1}_-(\al_1)=\nu(b)\frac{\ga(b(2\al_1-b))}{\ga(2b\al_1)}. \]
The resulting functional relation differs only very slightly from the 
Liouville case:  
\[ \frac{C(\al_4,\al_3,\al_1+b)}{C(\al_4,\al_3,\al_1)}=
\frac{(\nu(b))^{-1}\ga(b(2\al_1))\ga(b(2\al_1+b))\ga(b(\al_3+\al_4-\al_1-b))}
{\ga(b(\al_1+\al_3-\al_4))\ga(b(\al_1+\al_4-\al_3))\ga(b(\al_3+\al_4+\al_1-b))}
\]

\subsubsection{Case (1,2)}

The following remarkable fact facilitates the analysis considerably:
In the appendix it is shown that there exists a linear combination 
\begin{equation} 
\CG_{s-}^{\sst 1,2}(x,z)=a_{s-}\CF_{s-}^{\sst 1,2}(x,z)+a_{s\ti}
\CF_{s\ti}^{\sst 1,2}(x,z)
\end{equation} 
such that $\CG_{s+}^{\sst 1,2}\equiv\CF_{s+}^{\sst 1,2}$ and  
$\CG_{s-}^{\sst 1,2}(x,z)$ have the same monodromies as the (1,2)
Liouville conformal blocks 
$\CF_{s+}^{\sst L1,2}$ and  $\CF_{s-}^{\sst L1,2}$. Moreover, a second 
linear combination 
 \begin{equation} 
\CG_{s\ti}^{\sst 1,2}(x,z)=b_{s-}\CF_{s-}^{\sst 1,2}(x,z)+b_{s\ti}
\CF_{s\ti}^{\sst 1,2}(x,z)
\end{equation} 
has one dimensional monodromy representation.

In terms of the $\CG$'s it is therefore easy to write the most 
general crossing invariant and single valued combination of conformal blocks:
To simplify notation I will drop the superscript (1,2) and subscript s
in the following. 
\[ \Psi\equiv <\Phi^{j_4}\ldots\Phi^{j_1}> =\sum_{\si=+,-}
E^{\sst L}_{\si}\CG^{}_{\si}(x,z)\CG^{}_{\si}(\bx,\bz)
+E^{\sst D}_{}
\CG^{}_{s}(x,z)\CG^{}_{s}(\bx,\bz), \]
where $E^{\sst D}_{}$ is arbitrary.
Written in the $\CF$-basis of conformal blocks there 
appear non-diagonal terms:
\begin{equation}\label{offdiag} \begin{split}
\Psi =& E^{\sst L}_{+}\CF_{+}^{}\bar{\CF}^{}_{+} +
\bigl(E^{\sst L}_{-}a_{-}^2+E^{\sst D}_{}b_{-}^2\bigr)
\CF^{}_{-}\bar{\CF}^{}_{-}
     +\bigl(E^{\sst L}_{-}a_{\ti}^2+
            E^{\sst D}_{} b_{\ti}^2)\CF^{}_{\ti}
\bar{\CF}^{}_{\ti} \\
  & +
(E^{\sst L}_{-}a_-a_{\ti}+E^{\sst D}_{}b_{-}b_{\ti})
\Big(\CF^{}_{-}
\bar{\CF}^{}_{\ti}
+\CF_{\ti}^{}\bar{\CF}_{-}^{}\Big),
\end{split}\end{equation}
where the abbreviations $\CF_{\si}\equiv
\CF_{\si}(x,z)$, $\bar{\CF}_{\si}\equiv
\CF_{\si}(\bx,\bz)$ have been used.

However, it follows from the construction of conformal blocks given in 
\cite{T2} that only diagonal combinations of holomorphic and antiholomorphic
conformal blocks can appear in correlation functions of the \hp\ WZNW model.
Choose therefore $E^{\sst D}_{}$ such that the off-diagonal terms 
in \rf{offdiag} vanish.
With this choice one finds that 
\[ E_-=E_-^{\sst L}\frac{a_-}{b_{\ti}}(a_-b_{\ti}-b_-a_{\ti})
      =E_-^{\sst L}
\frac{\ga\big(b^{-1}(\al_1+\al_3+\al_4-\frac{3b^{-1}}{2})-1\big)
      \ga\left(2b^{-1}\al_1-1\right)}
     {\ga\big(b^{-1}(\al_1+\al_3+\al_4-\frac{b^{-1}}{2})-1\big)
      \ga\left(b^{-1}(2\al_1-b^{-1})-1\right)}
\]
from which it follows that
\begin{eqnarray*}
\frac{E_+}{E_-} & = & -\frac{
\ga\big(b^{-1}(\al_1+\al_3-\al_4-\frac{b^{-1}}{2})\big)
\ga\big(b^{-1}(\al_1-\al_3+\al_4-\frac{b^{-1}}{2})\big)}
{\ga\left(b^{-1}(2\al_1-b^{-1})\right)\ga\left(2b^{-1}\al_1-1\right)
\ga\big(b^{-1}(\al_3+\al_4-\al_1-\frac{b^{-1}}{2})\big)} \\
& & \ti\ga\big(b^{-1}(\al_1+\al_3+\al_4-\fr{b^{-1}}{2}-b)\big)
\end{eqnarray*}
Analogous to the Liouville case one finds 
\[ C^{\sst 1,2}_-(\al)=
\tilde{\nu}(b^{-1})\frac{\ga\bigl(b^{-1}(2\al-b)\bigr)}{\ga(2b^{-1}\al)} \]
leading to the functional equation
\[ \frac{
C(\al_4,\al_3,\al_1+b^{-1})}{C(\al_4,\al_3,\al_1)}=
\frac{(\tilde{\nu}(b^{-1}))^{-1}\ga(b^{-1}(2\al_1))\ga(b^{-1}(2\al_1+b^{-1}))
\ga(b^{-1}(\al_3+\al_4-\al_1-b^{-1}))}
{\ga(b^{-1}(\al_1+\al_3-\al_4))\ga(b^{-1}(\al_1+\al_4-\al_3))
\ga(b^{-1}(\al_3+\al_4+\al_1-b))}
\]
Comparing the functional equations found in the (2,1) and (1,2) cases
one observes that one is obtained from the other by $b\ra b^{-1}$. 
They can therefore be solved as in the Liouville case:
\begin{equation}\label{WZWthrpt}
 C(\al_1,\al_2,\al_3)=\frac{C_0(b)(\nu(b))^{-b^{-1}(\al_1+\al_2+\al_3)}\;
\up(2\al_1)\up(2\al_2)\up(2\al_3)}
{\up(\al_1+\al_2+\al_3-b)\up(\al_1+\al_2-\al_3)\up(\al_1+\al_3-\al_2)
\up(\al_2+\al_3-\al_1)} \end{equation}
where  $(\tilde{\nu}(b^{-1}))^b=(\nu(b))^{b^{-1}}$. 

\subsection{Reflection amplitude, two point function}

The amplitude $B(j)=R(j)$ can now be explicitely calculated from the three 
point function and the relation 
\begin{equation}
C\Bigl(
\begin{smallmatrix}j & j_2 & j_1 \\
                   x'& x_2 & x_1
\end{smallmatrix}\Bigr)
=-\pi\frac{\ga(j_1-j_2-j)\ga(j_2-j_1-j)}{\ga(-2j-1)}
\int_{\BC}d^2x'\; |x-x'|^{-4j-4}
C\Bigl(
\begin{smallmatrix}j & j_2 & j_1 \\
                   x'& x_2 & x_1
\end{smallmatrix}\Bigr)
\end{equation}
One finds
\begin{equation}\label{refamp}
R(j)=(\nu(b))^{-2j-1}\frac{\Ga(1-t^{-1}(2j+1))}{\Ga(1+t^{-1}(2j+1))}.
\end{equation}

\section{Fusion rules}

Fusion rules are the rules determining the set of irreducible 
representations contributing in the decomposition of the vector 
obtained by acting with a primary field $\Phi^{j_2}(x|z)$ on a primary state
$\Psi^{j_1}$. 

In most current algebra representations relevant in the
present context there are no nullvectors (cf. \cite{T2}).
The determination of fusion rules is therefore not an algebraic issue: 
In \cite{T2} it is shown that chiral vertex
operators between three representations with spins $j_3,j_2,j_1$ exist
for a certain range of {\it complex} values for the $j_i$ around the axis  
$j_i=-1/2+i\rho_i$ that furthermore allow 
meromorphic continuation to generic $j_i$.

Instead it will be argued that the issue of
fusion rules is intimately linked to the issue of spectral decomposition:
If $\Phi^{j_2}[v|\si)\Psi^{j_1}$ is a 
normalizable vector then it should be possible to expand it in terms of 
contributions from the representations $\CP_j$, $j=-1/2+i\rho$ constituting the
spectrum. In this case 
one will generically expect all representations appearing in the
spectrum to contribute, as is the case in the mini-superspace limit \cite{T1}.

From this point of view the problem is mainly to find  
criteria for the normalizability of the state $\Phi^{j_2}[v|\si)\Psi^{j_1}$.
The present section will present a 
heuristic argument based on the representation of primary fields and -states
in canonical quantization that will lead to
a precise conjecture on the fusion rules. 

\subsection{Spectral decomposition of $\CV$}

Motivated by the above mentioned result of Gawedzki and Kupiainen \cite{Ga},
I will assume that the space $\CV$ can be decomposed
into irreducible representations $\CP_j$: 
\begin{equation}
\CV=\int_{\rho>0}^{\oplus} d\rho \rho^2\;\,\CP_{-\frac{1}{2}+i\rho} .
\end{equation}
More explicitely I will assume that for a suitable subspace $\Phi\subset\CV$ 
of ``test-functions'' and its hermitian dual $\Phi^{\dagger}$ one has 
a set of maps 
\begin{equation}
\al_j:\Phi\ra \CP_j \qquad\mbox{and} \qquad \be_j:\CP_j\ra\Phi^{\dagger}
\end{equation}
that intertwine the $\hfg$ actions on $\Phi$ and $\CP_j$ resp.
$\CP_j$ and $\Phi^{\dagger}$, and allow to write the decomposition of elements
of $\Phi$ in the form 
\begin{equation}\label{absdec}
\Psi=-\frac{i}{4}
\int_{\frac{1}{2}+i\BR^+} dj \;\, \be_j\bigl(\al_j(\Psi)\bigr).
\end{equation}
The intertwining property for $\be_j$ in particular implies that eigenvectors
of $L_0+\bL_0$ are mapped to (generalized)
eigenvectors of $H$, so that expansion \rf{absdec} can be recast as an 
expansion into (generalized) eigenvectors of the Hamiltonian.

Consider therefore the eigenvalue condition $H\Psi_E=E\Psi_E$ for 
$\Psi\in\CF\ot\CC^{\infty}(H_3^+)$. It can be viewed as a system of 
differential
equations for the coefficients $\Psi_E^I(q,v,\bv)$ in the 
expansion
\begin{equation}\label{oscrep}
\Psi_E=\sum_{I}\CA_{I}\;\Om\;\Psi^{I}_E(q,v,\bv).
\end{equation}
The eigenvalue condition simplifies for $q\ra-\infty$, so one expects
$\Psi_e$ to be asymptotic to 
\begin{equation}\label{asym}
\Psi_E\sim \Psi_{\sst F,E} = 
e^{-2bjq}F^{+}_N(v,\bv)+
e^{-2b(-j-1)q}F^{-}_N(v,\bv)\bigr),
\end{equation}
where the $F^{\pm}_N(v,\bv)\in\CF_j$ have level $N$, 
$E=t^{-1}j(j+1)+N$.
Moreover, a qualitative analysis of the behavior for $q\ra\infty$ 
suggests that there are two linearly independent solutions with asymptotic 
behavior of the following type:
One horribly diverging (like $\sim \exp(e^{bq})$), the other rapidly 
converging (like $\sim \exp(-e^{bq})$). Suppressing the diverging solution
means that there must be a fixed relation $F^-_N=\CR(j)F^+_N$ between 
$F^{\pm}_N$, thereby defining an operator $\CR(j)$.
This is nothing but the statement that the interaction
term in the $H_3^+$-Hamiltonian acts perfectly reflecting as the 
``Liouville-wall'' does \cite{S}\cite{P}. 

These considerations lead to the conjecture that for any given value of
$j$ and vectors $F(v,\bv)\in\CF_j$
there exists a unique solution $\Psi_j[F]$ of $H\Psi_E=E\Psi_E$, 
$E=t^{-1}j(j+1)$ in $\CF\ot\CC^{\infty}(H_3^+)$
that has asymptotics \rf{asym} with $F^+\equiv F$, $F^-_N=\CR(j)F^+_N$
and which converges rapidly to zero for $q\ra\infty$. 
One has thereby found a representation of the map $\be_j$:
\begin{equation}
\be_j:\CF_j\ra\CF\ot\CC^{\infty}(H_3^+),\qquad 
\be_j(F)=\Psi_j[F].
\end{equation}
It is analogous to the construction of a harmonic function from its boundary
values. The 
intertwining property of the map $\Psi_j$ is nothing but the statement that
the generators $J^a_n$ and $\bJ^a_n$ go into the free field realizations
$I^a_n$, $\bI^a_n$ for $q\ra-\infty$. From this it also follows that 
the operator $\CR(j)$ must be the intertwining operator that 
establishes the equivalence between the Fock modules $\CF_j$ and 
$\CF_{-j-1}$. It is completely determined by the $\hfg$-intertwining property 
up to an overall factor $r(j)$ \cite{T2}. 
To unambigously define $r(j)$, 
consider the action of $\CR(j)$ on the level zero subspace of $\CF_j$: It must
be proportional to the $SL(2,\BC)$-intertwining operator 
$\CI_j$, so $r(j)$ will be
defined by
\begin{equation}\label{stref}
F^{-}_j(v,\bv)=r(j)\CI_j\bigl[F^{+}\bigr](v,\bv)\equiv 
r(j)\frac{2j+1}{\pi}
\int_{\BC}d^2v'\;|v-v'|^{4j}\; F^{+}_j(v',\bv').
\end{equation}
An observation that will be needed in the next section is that the 
intertwining operator $\CI_j$ is diagonalized by the Fourier transform
\begin{equation}
\tilde{F}(\mu,\bmu)=\frac{1}{2\pi}\int d^2v \;e^{\bmu \bx-\mu x}F(v,\bv).
\end{equation}
This follows from the fact that 
\begin{equation}
\CI_j\bigl[ e^{\mu x-\bmu\bx}\bigr] =(\mu\bmu)^{2j+1}
\frac{\Ga(+2j+1)}{\Ga(-2j-1)}e^{\mu x-\bmu\bx}.
\end{equation}
Let $\Psi^j_{\mu\bmu}$ denote the primary (generalized) eigenstate of $H$ that
has $q\ra -\infty$ asymptotics 
\begin{equation}\label{psidef}\begin{aligned}
\Psi^j_{\mu\bmu}\sim  e^{\mu v-\bmu\bv}\Bigl( 
:e^{-2bj\vf(\si)}:+ & S(j)(\mu\bmu)^{+2j+1}:e^{-2b(-j-1)\vf(\si)}:\Bigr),\\
 & S(j)=r(j)
\frac{\Ga(+2j+1)}{\Ga(-2j-1)}
\end{aligned}
\end{equation}
The state $\Psi^j_{\mu\bmu}$ thereby defined satisfies a simple reflection
property:
\begin{equation}
\Psi^j_{\mu\bmu}=S(j)(\mu\bmu)^{2j+1}\Psi^{-j-1}_{\mu\bmu}.
\end{equation}
It will be found in the next section that indeed the reflection amplitude 
$r(j)$ considered here is equal to the reflection amplitude $R(j)$ 
previously calculated in \rf{refamp}. 

\subsection{Primary fields}

The argument proposed here to find the fusion rules will require some 
qualitative information on how primary fields are represented in $\CV$.
It will be convenient to consider the Fourier transform of the operators 
$\Phi^{j}(x|\si)$:
\begin{equation}
\Phi^j_{\mu\bmu}(z)=\frac{1}{2\pi}(\mu\bmu)^{2j+1}
\int d^2x \;e^{\bmu\bx-\mu x}\;
\Phi^{j}(x|\si)
\end{equation}
In terms of the $\Phi^j_{\mu\bmu}(z)$ the primary field transformation law 
takes the following form:
\begin{equation}\label{fouprim}\begin{aligned}
{[}J_n^+,\Phi^j_{\mu\bmu}(z){]}=&
z^n\left(\bmu\frac{\pa^2}{\pa\bmu^2}-2j \frac{\pa}{\pa\bmu}\right)
\Phi^j_{\mu\bmu}(z) \\
{[}J_n^-,\Phi^j_{\mu\bmu}(z){]}=& -\bmu\Phi^j_{\mu\bmu}(z)
\end{aligned}\qquad {[}J_n^0,\Phi^j_{\mu\bmu}(z){]}=
z^n\left(\bmu\frac{\pa}{\pa\bmu}-j\right)\Phi^j_{\mu\bmu}(z).
\end{equation}
It is a difficult task to find such operators in terms of the elementary
operators used in canonical quantization. What can be found is again the 
$q\ra -\infty$ asymptotics of such an operator: In this asymptotics
one may replace $J^a_n$, $\bJ^a_n$ by $I^a_n$, $\bI^a_n$. 
The conditions obtained from \rf{fouprim} by that replacement are solved by
\begin{equation}
V^j_{\mu\bmu}(\si)\sim 
e^{\mu v(\si)-\bmu \bv(\si)}\Bigl( A^j_{\mu\bmu}
:e^{-2bj\vf(\si)}:+B^j_{\mu\bmu} (\mu\bmu)^{2j+1}:e^{-2b(-j-1)\vf(\si)}:
\Bigr),
\end{equation}
with a priori undetermined coefficients $A^j_{\mu\bmu}$, $B^j_{\mu\bmu}$. 
These are fixed by requiring that the operator $  V^j_{\mu\bmu}(z)$
corresponding to $V^j_{\mu\bmu}(\si)$ by (euclidean) time-evolution
creates the state $\Psi^j_{\mu\bmu}$ by the usual state-operator 
correspondence. For this one needs to consider the state $\Psi_0$
(the $SL(2,\BC)$-invariant ``vacuum'') that is defined by
$q\ra -\infty$ asymptotics $\Psi_0\sim \mbox{const.}$. 
Since this state transforms in the trivial representation of 
$\fg=\fsl\oplus\fsl$, the action of $V^j_{\mu\bmu}(\si)$ can only produce
states in the representation $\CP_j$. Considering the 
$q\ra -\infty$ asymptotics of $V^j_{\mu\bmu}(\si)\Psi_0$ yields
$A^j_{\mu\bmu}=1$, $B^j_{\mu\bmu}=S(j)(\mu\bmu)^{2j+1}$.
Comparing the reflection relation that 
$V^j_{\mu\bmu}(\si)$ satisfies with that of the state $\Psi^j_{\mu\bmu}$
finally allows to determine $r(j)$ resp. $S(j)$ as
\begin{equation}\label{Sdef}
S(j)=(\nu(b))^{2j+1}
\frac{\Ga(+2j+1)\Ga(1+t^{-1}(2j+1))}{\Ga(-2j-1)\Ga(1-t^{-1}(2j+1))}.
\end{equation}
It is no accident that 
the amplitude $S(j)$ appearing in \rf{psidef} is up to inessential factors
equal to the Liouville reflection amplitude discussed in \cite{ZZ}.

\subsection{Normalizability of fused state}

In view of the previous discussion it suffices to find out whether the 
vector $\Phi^{j_2}_{\mu_2\bmu_2}(\si)\Psi^{j_1}_{\mu_1\bmu_1}$ is in $\CV$ 
and can therefore
be expanded according to \rf{absdec}. 
The aim of the present subsection will be to find necessary and
(conjecturally) sufficient criteria for normalizability of the state
$\Psi_{21}\equiv\Phi^{j_2}_{\mu_2\bmu_2}(\si)\Psi^{j_1}_{\mu_1\bmu_1}$ 
by considering once more the $q\ra -\infty$ 
asymptotics of its representation in $\CF\ot\CC^{\infty}(H_3^+)$.
It is given by
\begin{equation}
\Psi_{21}\sim \; e^{(\mu_2+\mu_1)v-(\bmu_2+\bmu_1)\bv}
\sum_{s,s'=+,-}F_{s,s'}
\exp\Bigl(-2b\bigl(s\bigl(j_1+\fr{1}{2}\bigr)+
s'\bigl(j_2+\fr{1}{2}\bigr)-1\bigr)Q\Bigr)
\end{equation}
with $F_{s,s'}\in\CF$.
By using knowledge of the reflection amplitude $S(j)$
one finds as necessary condition for normalizability 
\begin{equation}\label{funran}
|\Re(j_1+j_2+1)|<\fr{1}{2}
\qquad\qquad |\Re(j_1-j_2)|<\fr{1}{2}.
\end{equation}
Note that the case that the $j_i$ correspond to representations from the 
spectrum, $j_i=-\frac{1}{2}+i\rho$ is well contained in that range, but
no case where $j_i=j_{r,s}$ is contained in it.

One is thereby lead to the conjecture that for $j_1$, $j_2$ satisfying
\rf{funran} the operator product expansion \rf{OPE} involves integration
over all $j$ with $j=-1/2+i\rho$, $\rho\in\BR$. 

In order to extend this conjecture on the fusion rules to general $j_2$, $j_1$
one should note that the coefficients appearing in the operator 
product expansion \rf{OPE} can be meromorphically continued to 
general complex $j_2$, $j_1$. In the process of analytic 
continuation it can happen that poles hit the contour of integration
$j=-1/2+i\rho_i$, $\rho\in\BR$. By deforming the contour one can always
rewrite the integral in terms of an integral over the original contour
plus a finite sum of residue contributions. This procedure allows to
unambigously define the fusion rules of two generic operators.

\subsection{Consistency of fusion rules with structure constants}

The fusion rules for the degenerate fields are severely restricted 
by the additional differential equations they satisfy. It is a nontrivial 
check on the previously obtained results on structure constants and
fusion rules that by analytic continuing the OPE from the range 
\rf{funran} to i.e. $j_1=j^{\pm}_{r,s}$ one indeed exactly recovers 
the restricted fusion rules of degenerate fields.

The basic mechanism is as follows:
If one sets i.e. $j_1$ in the expression for the structure constants 
\rf{WZWthrpt} to 
any of the values $j_{r,s}$ then the structure constants will generically 
vanish. This need not to be true for the residue terms picked up in
the analytic continuation from range \rf{funran} to
$j_1=j^{\pm}_{r,s}$. The integration in (\ref{ans})
therefore reduces to a sum over a finite number of terms only.  

Consider i.e. analytic continuation of $j_1$
to $j_{r,s}$ while keeping $j_2$ generic. 
Consider i.e. the poles 
of \rf{WZWthrpt} from $\al_1+\al_2+\al_3-b=-mb-nb^{-1}$ 
(Recall $\al_i=-bj_i$). 
The corresponding residue is proportional to
\[ \prod_{u=0}^{3}\left(
\prod_{i=0}^{m-1}\frac{1}{\ga(b(2\al_u+ib+nb^{-1}))}
\prod_{j=0}^{n} \frac{1}{\ga(b^{-1}(2\al_u+jb^{-1}))}
\right)
\]
It is easy to check that one needs $m\leq r$, $n\leq s-1$ to get non-vanishing
of these residue terms. Similarly one finds non-vanishing residues
\[ \mbox{from the poles at} \left\{\begin{array}{l}
\al_1+\al_2-\al_3=-mb-nb^{-1} \\ 
\al_2+\al_3-\al_1=(m+1)b+(n+1)b^{-1}\\
\al_1+\al_3-\al_2=-mb-nb^{-1} \end{array}\right\}
\mbox{only for}
\left\{\begin{array}{l} m\leq r, \quad n\leq s \\ 
m\leq r, \;\; n\leq s-1 \\
m\leq r, \;\; n\leq s 
\end{array}\right\} \]
The range of values for $j_3$ for which the residue terms may be non-vanishing
coincides with that given by the fusion rules \cite{T2}\cite{AY}
for fusion of degenerate with generic fields. 

Up to this point the choice of contour for the integration over $j$
(the fusion rules) did not enter the discussion at all.
The question that does crucially depend on the choice of contour is whether 
these residues are really picked up in the process of analytic continuation 
and contour deformation sketched above. 

To analyze this question it is 
convenient to start by considering the case that the analytic continuation 
has been performed such that $|\al_1-\al_2|<\frac{b}{2}$. Under that 
condition one picks up only poles of $\Upsilon^{-1}(\al_1+\al_2-\frac{b}{2}
-i\si)\Upsilon^{-1}(\al_1+\al_2-\frac{b}{2}+i\si)$. One should imagine 
$\al_1+\al_2$ to move slightly off the real axis in order to avoid that the
poles of these two $\Upsilon$-functions occur simultaneously. 

The condition that in analytically
continuing $\al_1$ to $-r\frac{b}{2}-s\frac{b^{-1}}{2}$ one has hit the 
pole with $\Re(\al_1+\al_2-\frac{b}{2})=-mb-nb^{-1}$ is
\begin{equation} \al_1+\al_2-\frac{b}{2}=-rb-sb^{-1}-\frac{1-\ep}{2}
<-mb-nb^{-1} 
\label{ineq}\end{equation}
if $\al_2=\al_1+b\frac{\ep}{2}$ for some $\ep\in (-1,1)$.
Integers $m,n$ such that $0\leq m\leq r$ and
$0\leq n\leq s$ will satisfy (\ref{ineq}) for {\it any} value of $b>0$. 
For any such $m$, $n$ one gets residue terms from the poles at both
$\al_1+\al_2+\al_3-b=-mb-nb^{-1}$ and  $\al_1+\al_2-\al_3-b=-mb-nb^{-1}$. 
In the special case $|\al_1-\al_2|<\frac{b}{2}$ considered one will therefore 
indeed pick up as many residue terms as are allowed by the fusion rules.
If one then considers more general cases for $|\al_1-\al_2|$ it is easy to
convince oneself that for any pole of 
$\Upsilon^{-1}(\al_1+\al_2-\frac{b}{2}
i\si)\Upsilon^{-1}(\al_1+\al_2-\frac{b}{2}+i\si)$ that is missed one gets 
an additional pole from 
$\Upsilon^{-1}(\al_1-\al_2+\frac{b}{2}
-i\si)\Upsilon^{-1}(\al_2-\al_1+\frac{b}{2}-i\si)$.

It follows that analytic continuation of the operator product expansion
\rf{OPE} indeed yields the correct fusion rules for the 
operator product expansion of a degenerate and a generic
field when the proposed fusion rules and structure constants are used. 

To see that also the coefficients are correctly reproduced it suffices 
to note that recursion relations for
the structure constants of degenerate fields can be derived exactly as
the functional relations for the 
structure constants of generic fields were derived. These 
recursion relations will coincide with those satisfied by the 
residues of $C(j_3,j_2,j_1)$.
Moreover, they allow to express the structure constants of any 
degenerate field $\Phi_{r,s}$ in terms of the $C_{\pm}^{1,2}(\al)$,
$C_{\pm}^{2,1}(\al)$ found above. Structure constants of degenerate
field and residues of structure constants of generic fields must therefore
coincide. 

\section{Conclusions}

Sufficient information has been obtained to define any n-point correlation
in the \hp-WZNW model on the sphere:

As discussed in more detail in \cite{T2}, one may characterize the 
conformal blocks as solutions of a KZ-type system of equations, which
here generically has continuous sets of solutions. There exist unique
power series solutions that can be identified with conformal blocks
corresponding to a fixed intermediate representation.

An explicit expression for the structure constants $C(j_3,j_2,j_1)$ was 
derived from the assumption that the degenerate fields are part of
the algebra of mutually local primary fields. 

Finally a discussion of the issue of fusion rules from the point of view
of canonical quantization was given.
Together with the pole structure of the 
$C(j_3,j_2,j_1)$ this led to a precise conjecture for the fusion rules for
general primary fields.

Clearly the task remains to prove crossing symmetry of the so defined
four-point functions resp. mutually locality of generic primary fields.
However, the assumptions that were made to derive structure
constants and fusion rules turned out to be consistent with each other
in a remarkable way: {\it The results of the previous subsection imply
that the proposed fusion rules are just right to make 
the conjecture of crossing symmetry of \rf{ans} for generic fields
compatible with the crossing symmetry in the case of one degenerate field,
which was the basis for the present derivation of the structure constants.}

To put the theory on firmer mathematical ground one may try to establish
and characterize duality transformations for the conformal blocks. 
The possibility of doing this by establishing relations to (noncompact)
quantum groups is currently under investigation. This may ultimately lead 
to a rigorous proof of the results on structure constants and fusion
rules discussed here.

From the point of view of physical applications let me note that one 
obtains almost immediately an important subset of the structure constants
for the euclidean black hole CFT from the \hp structure constants, 
namely those for which the winding number is conserved. 
One interesting consequence is already visible from the results presented
here: The exact reflection amplitude differs by certain quantum 
corrections from the corresponding quantity proposed in \cite{DVV}.
There it was obtained from a quantum mechanics which was proposed 
to describe tachyons in a black hole background. Quantum corrections
of the reflection amplitude indicate that the geometry probed by the
tachyons is not the geometry described by the classical sigma model metric.
It would be extremely interesting to see whether the resulting 
effective geometry can be efficiently described by some noncommutative
geometry as proposed in \cite{FG}

\newpage

\section{Appendix}

\subsection{Solutions to the Liouville decoupling equations}

In order to write down the solutions, introduce the parameters 
\[ \begin{array}{l@{\qquad}l} u=b(\al_1+\al_3+\al_4-3b/2)-1 & 
\bu=b^{-1}(\al_1+\al_3+\al_4-3b^{-1}/2)-1\\
v=b(\al_1+\al_3-\al_4-b/2) & \bv=b^{-1}(\al_1+\al_3-\al_4-b^{-1}/2)\\
w=b(2\al_1-b) & \bw=b^{-1}(2\al_1-b^{-1})
\end{array} \] 
Then the solutions to the 
decoupling equation \rf{Lioudec} that correspond to s-channel conformal blocks
are given by
\begin{eqnarray*}
\CF_{s+}^{\sst L2,1}&=&z^{b\al_1}(1-z)^{b\al_3}F(u,v,w;z) \\
\CF_{s-}^{\sst L2,1}&=&z^{1-b(\al_1-b)}(1-z)^{b\al_3}F(u-w+1,v-w+1,2-w;z),
\end{eqnarray*}
whereas the solutions for the t-channel are 
\begin{eqnarray*}
\CF_{t+}^{\sst L2,1}&=&z^{b\al_1}(1-z)^{b\al_3}F(u,v,u+v-w+1;1-z) \\
\CF_{t-}-^{\sst L2,1}&=&z^{b\al_1}(1-z)^{1-b(\al_3-b)}F(w-u,w-v,w-u-v+1;1-z).
\end{eqnarray*}
Similarly, the solutions in the (1,2)
case are obtained by replacing $b\ra b^{-1}$ 
and $u,v,w\ra \bu,\bv,\bw$.
\subsubsection{Fusion and braiding}

If a basis for the space of conformal blocks with diagonal 
monodromy around the origin is chosen then fusion 
matrix and the phases $\Om_{s\si}^{\sst L2,1}$, $\si=+,-$ defined by
\footnote{It will be assumed throughout that $z^{\la}$ is defined by
the principal value of the logarithm and that $\arg(z)\in (-\pi,0]$}
\[ \CF_{s\si}^{\sst L2,1}(e^{\pi i}z)=\Om_{s\si}^{\sst L2,1}
\CF_{s\si}^{\sst L2,1}(z) \]
form a complete set of duality data, i.e. completely determine the 
monodromies of conformal blocks.
Here the phases $\Om_{s\si}^{\sst L2,1}$ are read off as
\[ \Om_{s+}^{\sst L2,1}=e^{\pi i b\al_1} \qquad \Om_{s-}^{\sst L2,1}=
e^{\pi i(1-b(\al_1-b))} \]
The fusion matrix is calculated by using the standard results on analytic 
continuation of hypergeometric functions:
\begin{eqnarray*}
\CGL_+^{(0)} &=& \frac{\Ga(w)\Ga(w-u-v)}{\Ga(w-u)\Ga(w-v)}\CGL_+^{(1)}+
                \frac{\Ga(w)\Ga(u+v-w)}{\Ga(u)\Ga(v)}\CGL_-^{(1)} \\
\CGL_-^{(1)} &=& \frac{\Ga(2-w)\Ga(w-u-v)}{\Ga(1-u)\Ga(1-v)}\CGL_+^{(1)}-
                \frac{\Ga(2-w)\Ga(u+v-w)}{\Ga(u-w+1)\Ga(v-w+1)}\CGL_-^{(1)}
\end{eqnarray*}

\subsection{Solutions to decoupling equations (2,1)}

This case is of course well-known \cite{FZ,TK}. For the sake of a coherent 
presentation, I will nevertheless present the relevant results.

Equation $\pa_x\CF=0$ of course implies $\CF(x,z)=\CF^+(z)+x\CF^-(z)$. It 
straightforward to reduce the system of 
two ordinary differential equations that follows from the KZ-equations to 
hypergeometric equations for 
$\CF^+(z)$, $\CF^-(z)$. The normalization prescription is
\[ \CF(x,z)\sim z^{h_{21}-h_2-h_1}x^{j_1+j_2-j_{21}}(1+\CO(x)+\CO(z)) \]
in the limit of first taking $z\ra 0$, then $x\ra 0$. A set of normalized 
solutions for the s- and t-channel are then given by ($t_-=t^{-1}$)
\begin{align*} 
\CF_{s+}^{\sst 2,1} &=& z^{t_-j_1}(1-z)^{t_-j_3}\Bigl( & F(u+1,v,w;z)
                                    -x\frac{v}{w}F(u+1,v+1,w+1;z)\Bigr) \\ 
\CF_{s-}^{\sst 2,1} &=& z^{-t_-(j_1+1)}(1-z)^{t_-j_3} 
\Big( & xF(u-w+1,v-w+1,1-w;z)- \\
           &  &      &  z\frac{u-w+1}{1-w}F(u-w+2,v-w+1,2-w;z)\Big), 
\displaybreak[0] \\
\CF_{t+}^{\sst 2,1}&=& 
              (1-z)^{t_-j_3}z^{t_-j_1} \Big( & F(u+1,v,u+v-w+1;1-z)+ \\
          & &  &  (1-x)\frac{v}{u+v-w+1}F(u+1,v+1,u+v-w+2;1-z)\Big) \\
\CF_{t-}^{\sst 2,1} &=& 
(1-z)^{-t_-(j_3+1)}z^{t_-j_1}\Big( & (1-x)F(w-u,w-v,w-u-v;1-z)- \\
           & &    &  (1-z)\frac{w-v}{w-u-v}F(w-u,w-v+1,w-u-v+1;1-z)\Big).
\end{align*}

\subsubsection{Fusion and braiding}

The $\Om_{s\si}^{2,1}$-factors are 
\[ \Om_{s+}^{}=e^{\pi i t_-j_1} \qquad 
\Om_{-}^{\sst (0)}=e^{\pi i(1-t_-(j_1+1))} \]
The fusion relations now read
\begin{eqnarray*}
\CF_{s+}^{\sst 2,1} &=& \frac{\Ga(w)\Ga(w-u-v)}{\Ga(w-u)\Ga(w-v)}
                \CF_{t+}^{\sst 2,1}+
                \frac{\Ga(w)\Ga(u+v-w+1)}{\Ga(u+1)\Ga(v)}\CF_{t-}^{\sst 2,1} \\
\CF_{s-}^{\sst 2,1} &=& \frac{\Ga(1-w)\Ga(w-u-v)}{\Ga(-u)\Ga(1-v)}
                \CF_{t+}^{\sst 2,1}-
                \frac{\Ga(1-w)\Ga(u+v+w+1)}{\Ga(u-w+1)\Ga(v-w+1)}
                \CF_{t-}^{\sst 2,1}
\end{eqnarray*}

\subsubsection{Reduction to Liouville conformal blocks}

There are general arguments \cite{FGPP}\cite{PRY}
that taking the limit $x\ra z$ relates the WZNW conformal blocks to 
their Liouville counterparts. Here one finds
\begin{equation}
\CF_{s+}^{\sst 2,1}(z,z)= \CF_{s+}^{\sst L2,1}(z) \qquad\qquad
\CF_{s-}^{\sst 2,1}(z,z)=\frac{u}{u-w+1}\CF_{s-}^{\sst L2,1}(z) 
\end{equation}

\subsection{Solutions to decoupling equations (1,2)}

It is easy to see that the decoupling equation \rf{12eqn} for the (1,2)
degenerate field 
coincides with the third order ordinary differential 
equation satisfied by the generalized hypergeometric function 
$F_1(\al,\be,\be',\ga;x,z)$ of Appell and 
Kamp\'{e} de Feriet (\cite{AK}, Chap. III, eqn. (31)), provided one 
identifies the parameters as follows:
\begin{equation}
\begin{array}{cc} 
\al=-\D=j_4-j_1-j_3+t/2 & \be=t\\
\be'=\D-1+t-2j_1-2j_3=t/2-j_1-j_3-j_4-1 & \ga=t-2j_1
\end{array}
\end{equation}
Equation (II) has three linearly independent solutions, 
unique up to linear combinations with
arbitrary functions of $z$ as coefficients. 
The $z$ dependence is determined by the KZ-equation: If one sets
\begin{equation}\label{ffrel} \CF(x,z)=z^{-j_1}(1-z)^{-j_3}F(x,z), 
\end{equation}
then the KZ equation for $\CF$ is equivalent to a similar equation 
for $F$ which may easily 
seen to be one of the partial differential equations satisfied by $F_1$, 
namely the first of eqns. (13), 
Chap. III in \cite{AK}. The function $F$ can therefore be any linear 
combination of the three linearly
independent solutions of the system of partial differential equations 
for $F_1$. As shown in \cite{AK}, 
Chap. III, pp. 55-65, each of them can be expressed in terms of the 
function $F_1$ itself, for which one
has representations 
\begin{equation}\label{intrep}
 F_1(\al,\be,\be',\ga;x,z)=\frac{\Ga(\ga)}{\Ga(\al)\Ga(\ga-\al)}
\int_0^1 dt\; t^{\al-1}(1-t)^{\ga-\al-1}
(1-xt)^{-\be}(1-zt)^{-\be'}
\end{equation}
or as the power series
\[ F_1(\al,\be,\be',\ga;x,z)=\sum_{n,m=0}^{\infty}
\frac{(\al)_{m+n}(\be)_m(\be')_n}{\ga_{m+n}}
\frac{x^m}{m!}\frac{y^n}{n!} \]
The task now is to identify solutions to the $F_1$ equations with 
conformal blocks corresponding to 
definite intermediate representations. The s-channel conformal blocks 
should be of the form \cite{T2}
\begin{equation}\label{solform} \CF_s=\sum_{n=0}^{\infty}z^{s+n}f_n(x), \quad
f_0(x)=\sum_{m=0}x^{r+m}g_n \end{equation}
It may be shown that for generic (i.e. non-integer) values of $2j_1-t$ and $t$ 
there indeed exist three linearly independent solutions of the form 
\rf{solform}. These are uniquely 
specified up to multiplication by $(x,z)$-independent factors once one 
has chosen one of the three 
possible values for $r$:
\begin{equation}\label{rval}
r_+=0\qquad r_-=-t \qquad r_{\ti}=2j_1+1-t.
\end{equation}
The values of $s$ for the corresponding solutions are 
\begin{equation}\label{sval}
s_+=-j_1\qquad s_-=j_1+1 \qquad s_{\ti}=-j_1.
\end{equation}
In order to identify these solutions with conformal blocks one needs to
have $r_i=j_1+j_2-j_{21,i}$ ($i\in\{+,-,\ti\}$) and 
$s_i=h_{j_{21,i}}-h_{j_2}-h_{j_1}$ which is satisfied by
\begin{equation}\label{jval}
j_{21,+}=j_1-t/2 \qquad j_{21,-}=j_1+t/2 \qquad j_{21,\ti}=-j_1-1+t/2.
\end{equation}
One thereby recovers a special case of the fusion rules derived in \cite{AY}. 
Note that the values $s_+$ and $s_{\ti}$ coincide.

The task is now to find the explicit expressions for these solutions in 
terms  of the $F_1$-functions.
The relevant solutions of the $F_1$-system will be denoted 
$F_{s+}$, $F_{s-}$,  $F_{s\ti}$ respectively, the 
corresponding conformal blocks by $F\ra\CF$, cf. \rf{ffrel}.
A table of solutions to the system of equations satisfied by $F_1$ that 
possess simple integral 
representations similar to \rf{intrep} has been given in \cite{AK}, 
Chap. III, Sect. XV. The ones that 
will be needed below are
\begin{eqnarray*}
Z_1 &=& F_1(\al,\be,\be',\ga;x,z)\\
Z_2 &=& F_1(\al,\be,\be',\al+\be+\be'+1-\ga;1-x,1-z)\\
Z_5 &=& z^{\be+1-\ga}(1-z)^{\ga-\al-1}(x-z)^{-\be}
        F_1\bigl(1-\be',\be,\al+1-\ga,2+\be-\ga,\fr{z}{z-x},\fr{z}{z-1}
        \bigr)\\
Z_7 &=& z^{\be+\be'-\ga}(1-z)^{\ga-\al-\be'}(x-z)^{-\be}
        F_1\bigl(1-\be',\be,\ga-\be-\be',\ga+1-\al-\be';\fr{z-1}{z-x}, 
        \fr{z-1}{z}\bigr)\\
Z_8 &=& x^{-\al}F_1\bigl(\al,\al+1-\ga,\be',\al+1-\be;\fr{1}{x},       
        \fr{z}{x}\bigr)
\end{eqnarray*}
In view of 
above considerations it suffices to first take the limit $z\ra 0$ 
(our variable $z$ corresponds to $y$
in loc. cit.), before taking $x\ra 0$ in order to compare the asymptotic 
behavior of the solutions
given in loc. cit., p. 62 with that expected for the conformal blocks. 
In this way one easily
identifies $F_{s+}=Z_1$ and $F_{s-}=Z_5$.
Finding the solution corresponding to $F_{s\ti}$ is a little more 
complicated since none of the 
solutions given in loc. cit. has the required asymptotic behavior. 
Consider however 
\[ Z_8=x^{-\al}F_1\left(\al,\al+1-\ga,\be',\al+1-\be;\frac{1}{x},
\frac{z}{x}\right) \]
From the power series expansion of $F_1$ one easily sees that $Z_8$ 
is analytic as function of $z$ in
a neighborhood of $z=0$ and 
\[ Z_8|_{z=0}=x^{-\al}F(\al,\al+1-\ga;\al+1-\be;1/x),  \]
where $F(\al,\be,\ga,x)$ is the ordinary hypergeometric function. 
This is rewritten in terms of 
functions with simple asymptotics for $x\ra 0$ by using standard
 results on the analytic 
continuation of hypergeometric functions:  
\begin{eqnarray*} 
Z_8|_{z=0}& = & e^{\pi i\al}
   \frac{\Ga(\al+1-\be)\Ga(1-\ga)}{\Ga(\al+1-\ga)\Ga(1-\be)}F(\al,\be;\ga;z)\\
          & + & e^{\pi i(\ga-\al-1)}
\frac{\Ga(\al+1-\be)\Ga(\ga-1)}{\Ga(\al)\Ga(\ga-\be)}
                 x^{1-\ga}F(\al+1-\ga,\be+1-\ga;2-\ga;x)
\end{eqnarray*}
for $\arg(-1/x)\in (-\pi,0]$. 
Since $1-\ga=2j_1+1-t$, the second term has the asymptotics required for 
$F_{\ti}$ which may 
therefore be represented as 
\[ F_{s\ti}=e^{-\pi i (\al+1-\ga)}
\frac{\Ga(\al)\Ga(\ga-\be)}{\Ga(\al+1-\be)\Ga(\ga-1)}
(Z_8-e^{\pi i\al}Z_1)\qquad
\mbox{for }\arg(-1/x)\in (-\pi,0], \]
with a similar expression for $\arg(-1/x)\in (0,\pi]$. 
In this way one finds the following two bases for solutions:
\begin{align*}
F_{s+}=& Z_1 \\
F_{s\ti} =& e^{\pi i(\al+1-\ga)}
                  \frac{\Ga(\al)\Ga(\ga-\be)}{\Ga(\al+1-\be)\Ga(\ga-1)}
                  \left( Z_8-e^{-\pi i \al}\frac{\Ga(\al+1-\be)
                        \Ga(1-\ga)}{\Ga(\al+1-\ga)\Ga(1-\be)}Z_1 \right)\\
F_{s-}=& Z_5 \displaybreak[0] \\
F_{t+}=& Z_2 \\
F_{t\ti} =& \frac{e^{\pi i(\al+\be+\be'-\ga)}\Ga(\al)
                 \Ga(1+\al+\be'-\ga)}{\Ga(\al+1-\be)\Ga(\al+\be+\be'-\ga)}
                  \left( Z_8-\frac{\Ga(\al+1-\be)\Ga(\ga-\al-\be-\be')}
                  {\Ga(1-\be)\Ga(\ga-\be-\be')}Z_2\right)\\
F_{s-}=& Z_5
\end{align*}

\subsubsection{Fusion and braiding}

The $\Om_{s\si}^{\sst 1,2}$-factors are 
\[ \Om_{s+}^{\sst 1,2}=e^{-\pi ij_1} \qquad 
\Om_{s-}^{\sst 1,2}=e^{\pi i(j_1+1)}
\qquad 
\Om_{s\ti}^{\sst 1,2}=e^{-\pi ij_1} \]
The fusion matrix may be computed from the integral representations of 
the $Z_i$-functions
by using the technique exemplified in chapter III, section XVI of \cite{AK}. 
The result is
\begin{align*}
\CF_{s+}^{\sst 1,2} =& 
   \frac{\Ga(\ga)\Ga(\ga-\be-\be'-\al)}{\Ga(\ga-\al)\Ga(\ga-\be-\be')}
   \CF_{t+}^{\sst 1,2}
  +\frac{\Ga(\ga)\Ga(\al+\be'-\ga)}{\Ga(\al)\Ga(\be')}
   \CF_{t-}^{\sst 1,2}\\
  & +\frac{\Ga(\ga)\Ga(\ga-\al-\be')\Ga(\al+\be+\be'-\ga)}{\Ga(\al)
            \Ga(\ga-\al)\Ga(\be)} \CF_{t\ti}^{\sst 1,2} \displaybreak[0] \\
\CF_{s-}^{\sst 1,2} =& 
    \frac{\Ga(2+\be-\ga)\Ga(\ga-\be-\be'-\al)}{\Ga(1-\be')\Ga(1-\al)}
    \CF_{t+}^{\sst 1,2}
   +e^{\pi i \be}\frac{\Ga(2+\be-\ga)\Ga(\al+\be'-\ga)}{\Ga(1+\be+\be'-\ga)
    \Ga(1-\ga+\al)}\CF_{t-}^{\sst 1,2}\\
           & +e^{\pi i (\be+\be'+\al-\ga)}\frac{\Ga(2+\be-\ga)
                   \Ga(\ga-\al-\be')\Ga(\al+\be+\be'-\ga)}
              {\Ga(1-\be')\Ga(1+\be+\be'-\ga)\Ga(\be)}
            \CF_{t\ti}^{\sst 1,2} \displaybreak[0] \\
\CF_{s\ti}^{\sst 1,2}=& \frac{\Ga(\ga-\be)}{\Ga(1-\be)}\Bigg(
          \frac{\Ga(2-\ga)\Ga(\ga-\be-\be'-\al)}{\Ga(\ga-\be-\be')\Ga(1-\al)}
          \CF_{t+}^{\sst 1,2}
     -e^{\pi i \ga}\frac{\Ga(2-\ga)\Ga(\al+\be'-\ga)}{\Ga(1+\al-\ga)\Ga(\be')}
                 \CF_{t-}^{\sst 1,2}\\ 
              & +e^{\pi i \ga}\left( e^{\pi i (\be+\be'-\ga)}
                 \frac{\sin\pi\ga}{\sin\pi\be}
                 -\frac{\sin\pi(\ga-\al)}{\sin\pi(\ga-\al-\be')}\right)
                 \frac{\Ga(2-\ga)\Ga(\al+\be+\be'-\ga)}{\Ga(\be)
             \Ga(1+\al+\be'-\ga)}\CF_{t\ti}^{\sst 1,2}\Bigg)
\end{align*}

\subsubsection{Alternative choice of basis for conformal blocks}

An important simplification of the fusion relations is achieved by using the 
following linear combinations of the conformal blocks:
\begin{align*} 
\CG_{s+}^{\sst 1,2}=& \CF_{s+}^{\sst 1,2}\\
\CG_{s-}^{\sst 1,2}=& \frac{\Ga(\be+\be')\Ga(1+\be-\ga)}{\Ga(\be)
\Ga(\be+\be'+1-\ga)}\CF_{s\ti}^{\sst 1,2}
 +\frac{\Ga(\be+\be')\Ga(\ga-\be-1)}{\Ga(\ga-1)\Ga(\be')}\CF_{s-}^{\sst 1,2}\\
\CG_{s\ti}^{\sst 1,2}=& 
  \frac{\Ga(1-\be)\Ga(1+\be-\ga)}{\Ga(2-\ga)}\CF_{s\ti}^{\sst 1,2}
 +\frac{\Ga(1-\be')\Ga(\ga-\be-1)}{\Ga(\ga-\be-\be')}\CF_{s-}^{\sst 1,2}
 \displaybreak[0] \\
\CG_{t+}^{\sst 1,2}=& \CF_{t+}^{\sst 1,2}\\
\CG_{t-}^{\sst 1,2}=& \frac{\Ga(\be+\be')\Ga(\ga-\al-\be')}{\Ga(\ga-\al)
 \Ga(\be)}\CF_{t\ti}^{\sst 1,2}
 +\frac{\Ga(\be+\be')\Ga(\al+\be'-\ga)}{\Ga(\al+\be+\be'-\ga)\Ga(\be')}
                      \CF_{t-}^{\sst 1,2}\\
\CG_{t\ti}^{\sst 1,2}=& 
  \frac{\Ga(1-\be)\Ga(\ga-\al-\be')}{\Ga(1+\ga-\al-\be-\be')}
  \CF_{t\ti}^{\sst 1,2}
  +\frac{\Ga(1-\be')\Ga(\al+\be'-\ga)}{\Ga(1+\al-\ga)}\CF_{t-}^{\sst 1,2}
\end{align*}
The fusion relations for this basis read
\begin{align*}
\CG_{s+}^{\sst 1,2} =& 
  \frac{\Ga(\ga)\Ga(\ga-\be-\be'-\al)}{\Ga(\ga-\al)\Ga(\ga-\be-\be')}
  \CG_{t+}^{\sst 1,2}
 +\frac{\Ga(\ga)\Ga(\al+\be+\be'-\ga)}{\Ga(\be+\be')\Ga(\al)}
  \CG_{t-}^{\sst 1,2}\\
  \CG_{s-}^{\sst 1,2} =& \frac{\Ga(2-\ga)\Ga(\ga-\be-\be'-\al)}{\Ga(1-\al)
  \Ga(1-\be-\be')}\CG_{t+}^{\sst 1,2}
  +\frac{\Ga(2-\ga)\Ga(\al+\be+\be'-\ga)}{\Ga(1+\al-\ga)\Ga(1+\be+\be'-\ga)}
  \CG_{t-}^{\sst 1,2}\\
\CG_{s\ti}^{\sst 1,2} =& e^{\pi i(\be+\be'+1)}\CG_{t\ti}^{\sst 1,2}
\end{align*}
Two remarks will be important:
\begin{enumerate}
\item The fusion matrix in the subspace spanned by 
$\CG_{s+}^{\sst 1,2}$ $\CG_{s}-^{\sst 1,2}$ is identical to 
that of the Liouville conformal blocks 
$\CF_{s+}^{\sst L1,2}$, $\CF_{s-}^{\sst 1,2}$. 
This will be further explained in the next subsection.
\item The conformal block $\CG_{s\ti}^{\sst 1,2}$ 
transforms into itself under fusion.
\end{enumerate}

\subsubsection{Reduction to Liouville conformal blocks}

The analysis of the behavior of the $G$-basis for the conformal blocks is 
facilitated by the observation that 
\begin{eqnarray*}
\CG_{s-}^{\sst 1,2} &=& 
z^{-j_1}(1-z)^{-j_3}\frac{\Ga(\al)\Ga(\be+\be')}{\Ga(\ga-1)\Ga(\be')}\left(Z_2-
    \frac{\Ga(\al+\be+\be'-\ga)\Ga(1-\ga)}{\Ga(\be+\be'+1-\ga)\Ga(\al+1-\ga)}
    Z_1\right) \\
\CG_{t-}^{\sst 1,2} &=& 
     z^{-j_1}(1-z)^{-j_3}\frac{\Ga(\al)\Ga(\be+\be')}{\Ga(\ga)
  \Ga(\al+\be+\be'-\ga)}\left(Z_1-
              \frac{\Ga(\ga)\Ga(\ga-\al-\be-\be')}{\Ga(\ga-\al)
  \Ga(\ga-\be-\be')}Z_2\right)
\end{eqnarray*}
Since $F_1(\al,\be,\be',\ga;x,z)$ is analytic in $x$ around $x=z$ it follows 
that
$\CG_{s+}^{\sst 1,2}$, $\CG_{s-}^{\sst 1,2}$, $\CG_{t+}^{\sst 1,2}$, 
$\CG_{t-}^{\sst 1,2}$ all share that property. 
By using 
\[  F_1(\al,\be,\be',\ga;z,z)=F(\al,\be+\be',\ga;z), \]
standard relations on analytic continuation of hypergeometric functions
and observing that
\[ \al=\bu \qquad \be+\be'=\bv \qquad \ga=\bw \]
one finds that $\CG_{s+}^{\sst 1,2}$, $\CG_{s-}^{\sst 1,2}$, 
$\CG_{t+}^{\sst 1,2}$, $\CG_{t-}^{\sst 1,2}$ reduce to 
$\CF_{s+}^{\sst L1,2}$, $\CF_{s-}^{\sst L1,2}$, $\CF_{t+}^{\sst L1,2}$, 
$\CF_{t-}^{\sst L1,2}$ respectively for $x\ra z$.
$\CG_{\ti}$ 
however will not be analytic in $x$ around $x=z$ but rather behave as 
\[ \CG_s\sim (x-z)^{1-\be-\be'}(C_1+\CO(x-z)) \]

\newcommand{\CMP}[3]{{\it Comm. Math. Phys. }{\bf #1} (#2) #3}
\newcommand{\LMP}[3]{{\it Lett. Math. Phys. }{\bf #1} (#2) #3}
\newcommand{\IMP}[3]{{\it Int. J. Mod. Phys. }{\bf A#1} (#2) #3}
\newcommand{\NP}[3]{{\it Nucl. Phys. }{\bf B#1} (#2) #3}
\newcommand{\PL}[3]{{\it Phys. Lett. }{\bf B#1} (#2) #3}
\newcommand{\MPL}[3]{{\it Mod. Phys. Lett. }{\bf A#1} (#2) #3}
\newcommand{\PRL}[3]{{\it Phys. Rev. Lett. }{\bf #1} (#2) #3}
\newcommand{\AP}[3]{{\it Ann. Phys. (N.Y.) }{\bf #1} (#2) #3}
\newcommand{\LMJ}[3]{{\it Leningrad Math. J. }{\bf #1} (#2) #3}
\newcommand{\FAA}[3]{{\it Funct. Anal. Appl. }{\bf #1} (#2) #3}
\newcommand{\PTPS}[3]{{\it Progr. Theor. Phys. Suppl. }{\bf #1} (#2) #3}
\newcommand{\LMN}[3]{{\it Lecture Notes in Mathematics }{\bf #1} (#2) #2}

\end{document}